\documentclass[a4paper,twocolumn,useAMS,usenatbib]{mn2e}
\usepackage[totalwidth=480pt, totalheight=680pt]{geometry}
\usepackage{graphicx}
\usepackage{aas_macros}
\usepackage{amssymb}   

\newcommand{\rg}{MRC B0319$-$454 }
\newcommand{\rgns}{MRC B0319$-$454}
\newcommand{\msolar}{M_\odot}

\title[\rgns: Probing large-scale structure]{\rgns: Probing the large-scale structure with a
giant radio galaxy}
\author[Safouris, Subrahmanyan, Bicknell \& Saripalli]
{V. Safouris$^{1,2}$\thanks{E-mail: vicky@mso.anu.edu.au (VS)},
  R. Subrahmanyan$^{2,3}$, 
G. V. Bicknell$^{1}$ and L. Saripalli$^{2,3}$  
\\
$^{1}$Research School of Astronomy and Astrophysics, Mount Stromlo Observatory, 
Australian National University,\\
 Cotter Road, Weston ACT 2611, Australia
\\
$^{2}$Australia Telescope National Facility, CSIRO, Locked Bag 194, 
Narrabri, NSW 2390, Australia
\\
$^{3}$Raman Research Institute, C V Raman Avenue, Sadashivanagar, 
Bangalore 560080, India}

\begin{document}


\pagerange{\pageref{firstpage}--\pageref{lastpage}} \pubyear{2002}

\maketitle

\label{firstpage}

\begin{abstract}
We present an investigation of the relationships between the
radio properties of a giant radio galaxy MRC B0319-454 and the surrounding
galaxy distribution with the aim of examining the influence of
intergalactic gas and gravity associated with the large-scale structure on
the evolution in the radio morphology.  Our new radio continuum
observations of the radio source, with high surface brightness
sensitivity, images the asymmetries in the megaparsec-scale radio
structure in total intensity and polarization.  We compare these with the 3-D
galaxy distribution derived from galaxy redshift surveys.  Galaxy density
gradients are observed along and perpendicular to the radio axis: the
large-scale structure is consistent with a model wherein the galaxies
trace the ambient intergalactic gas and the evolution of the radio
structures are ram-pressure limited by this associated gas.  Additionally,
we have modeled the off-axis evolution of the south-west radio lobe as
deflection of a buoyant jet backflow by a transverse gravitational field:
the model is plausible if entrainment is small.  The case study presented
here is a demonstration that giant radio galaxies may be useful probes of
the warm-hot intergalactic medium believed to be associated with
moderately over dense galaxy distributions.

\end{abstract}


\section{Introduction}

In a powerful radio galaxy, reaccelerated jet material inflates the
synchrotron lobes after passage through the termination shocks at the jet
ends \citep{scheuer74,begelman84}. The synchrotron lobes interact 
with the surrounding thermal gas and their morphologies are a result 
of this interaction. In cluster environments, where XMM-Newton and 
Chandra have been able to detect and image the relatively dense 
intra-cluster medium (ICM) in X-rays, the total intensity radio and
X-ray contours follow each other and X-ray holes are observed
at the locations of the radio lobes
\citep[e.g.][]{boehringer93,mcnamara00,nulsen02,birzan04}. 
These holes are evidence that the expanding radio lobes interact
with, and displace the X-ray emitting gas. 
Lobe interaction with intra-cluster gas is also apparent in the
bending of radio plumes in wide-angle and narrow-angle tailed radio
sources \citep[e.g.][]{bliton98,hardcastle05,douglass08}. 
The morphologies of these sources suggest that the radio
tails are deflected behind the moving host galaxy, due to the
ram-pressure exerted by the ICM.
\par
Giant radio galaxies, which have linear sizes
of $>1$~Mpc, tend to reside outside of rich clusters \citep{jamrozy04}. 
Their giant lobes, which extend well beyond the  interstellar media 
(ISM) and coronal halos of  their host galaxies,
represent the interaction between the light synchrotron lobe plasma 
and the heavier intergalactic medium (IGM). 
If galaxies are a reliable tracer of the IGM gas, then we expect the
distribution of gas around a giant radio galaxy to depend on the
position of the host galaxy relative to large-scale structure
in the IGM.  
Since the evolution of the radio lobes depends on their interaction with 
this ambient gas, large scale structure (and the associated gas) 
is also expected to determine the lobe dynamics 
in giant radio sources. 
For example, if the parent optical galaxy resides at a boundary between an
over-density and a void, then there may be a gradient in the gas density
surrounding the radio source. If this gradient is aligned with the radio jet
axis, then we would expect an asymmetry to arise in the lobe lengths,
due to differing ram-pressure limitations in the ambient gas on the two sides.
If, on the other hand, the gradient in the gas density is transverse to the
jet axis, then we would expect the light synchrotron lobes to evolve
transverse to the jet axis and in the direction of decreasing ambient
density and pressure due to buoyancy. 
\par
The diffuse gas in the environs of giant radio galaxies, outside of rich clusters, 
most likely pertains to the warm-hot phase of the IGM, whose existence at 
low-redshifts has been
predicted by large-scale cosmological hydrodynamical simulations of galaxy
formation \citep{cen99, cen06, dave01}. In the simulations, the warm-hot gas
follows the filamentary galaxy distribution on large scales and represents
unvirialized over-densities in the range 10-30. Little is known about this
warm-hot gas since  its thermal emission is not detectable by present day
X-ray telescopes. However, the shape taken by the lobes of a giant radio
galaxy is a visible manifestation of their interaction with 
the surrounding gas. These galaxies, therefore, make ideal probes of the
unseen ambient medium, which may be the warm-hot intergalactic 
medium (WHIM).
\par
Recently, the 1.8~Mpc lobes of the giant radio galaxy MSH 05$-$2{\it 2} were
used to constrain the properties of its ambient thermal gas \citep{subrahmanyan08}. The
lobes in this galaxy appear to be relicts and there is an observed 
asymmetry, which appears to be related to an anisotropy in
the local galaxy distribution. A comparison of the properties of the radio
lobes and those of the ambient gas, which were derived from the surrounding
galaxy distribution, indicated that the lobes were highly overpressured
despite their relict appearance. Alternatively, the density-temperature product for the
ambient IGM might be an order of magnitude larger than that predicted by
structure formation models indicating significant feedback in the IGM.
\par 
In this paper, we present a similar study of another giant radio galaxy
\rg (also referred to previously as MSH 03$-$4{\it 3} and PMN
J0321$-$4510) and its environment. The 2.5 Mpc radio source
is located within a galaxy filament of the Horologium-Reticulum super-cluster 
 \citep{fleenor05} at a redshift of $z\approx0.6$. Its 26\arcmin size on the sky, 
make the measurement of the surrounding galaxy distribution
possible with multi-object fibre instruments such as AAOmega on the
Anglo-Australian Telescope. 
\rgns, therefore, provides a further opportunity to
study the morphology of a giant radio galaxy
and its relationship to the surrounding galaxy distribution.
Such studies are a first step in using giant radio galaxies as
probes of the unseen intergalactic gas associated with large-scale
galaxy structures.

\par
The giant radio source \rg 
was previously observed at 843-MHz with the Molonglo Observatory Synthesis
Telescope (MOST), and with the Australia Telescope Compact Array (ATCA) at 20, 13 and
6 cm \citep{jones89, saripalli94}. In the latter work, Saripalli et al.
presented a detailed study of the morphology of the giant radio galaxy
pointing out a number of features that make the radio source unusual: a unique
configuration of five compact hot spots in one of the lobes, 
a prominent jet and counter-jet detected out to exceptional distances, 
and lobes that are asymmetrically shaped and positioned about the core.
Considering the likely expansion velocity of the source, it was argued that 
light travel-time effects
were not responsible for the asymmetry in the lobe 
separations from the core. Based on an examination of the projected  galaxy distribution 
around the host, Saripalli et al. attributed this feature
to corresponding asymmetries in the ambient intergalactic medium.

\par
Given the Mpc size and pronounced asymmetries in the radio morphology,
\rg is an excellent example of a radio galaxy where the morphology is
shaped by the IGM. 
In this paper,
we examine the interaction between the lobes
and the ambient IGM using our new radio observations made with
the specific purpose of imaging the radio lobes fully.  Our new improved radio
images of \rg have been made using an observing mode designed to have
high surface brightness sensitivity and in full polarization. We use these
in conjunction with redshift measurements of the surrounding galaxy distribution, 
in an attempt to model the interaction between the giant lobes with 
their IGM environment. Our radio observations include polarization 
measurements for the first time.
\par
The paper is organized as follows: In Sect.~\ref{s:radio} we present our new
radio continuum images of \rgns. In Section 3, we describe the galaxy redshift
data; in the sections that follow an attempt is made towards understanding
the lobe-IGM interaction assuming that the galaxy distribution traces the gas
distribution on large scales. 
Throughout we adopt a flat cosmology with parameters $\Omega_{\rm 0} = 0.3$, $\Omega_
{\Lambda }= 0.7$ and a Hubble constant $H_{0}=71$ km s$^{-1}$ Mpc$^{-1}$. The 
host elliptical galaxy,  ESO~248-G10 (also cataloged as AM~0319$-$452),  has an R-band absolute
magnitude $M_R = -23.7$  \citep{saripalli94}. Previously, the redshift of the
host was estimated to be $z=0.0633$ \citep{jones89}; our new data, presented
herein, give an estimate of $z = 0.0622$.  At this redshift, 1\arcsec = 1.2 kpc.

\section{Radio continuum imaging}
\label{s:radio}

New radio continuum observations of \rg were made with the ATCA during the
period 2003 September to 2004 April. These were aimed at accurately   
imaging the extended total intensity and polarization structures in the
26\arcmin~source in the 20 and 13~cm bands. Earth-rotation synthesis
with full UV coverage were made in 5 separate array configurations 
that emphasized low spatial frequencies: 1.5D and 1.5A 1.5-km
arrays, a 750B 750-m array, a EW352 352-m array, and a
compact EW214 214-m array. The relatively compact configurations 
were for improving sensitivity to the low
surface brightness structures, while the relatively longer 1.5-km arrays were to enable
imaging with sub-arcminute resolution. Visibilities were
measured simultaneously in two bands, 128-MHz wide, centered at 1378
and 2368~MHz. The continuum bands were covered in 13 independent channels.  A
journal of the observations is in Table~\ref{t:journal}.  
The ATCA antennas have primary beam full width half 
maximum (FWHM) of about 35\arcmin~ and
21\arcmin~ respectively in the 20 and 13~cm bands.  Therefore, to accurately
image the extended emission in this 26\arcmin source 
we adopted the approach of mosaic observing and covered the double radio
source with 8 separate pointings in a  $4\times2$ grid.
A grid spacing of about 10\arcmin, which is approximately equal to
the sky plane Nyquist sampling requirement for observing
in the 13~cm band, was adopted for the pointings.
\par
Data at both frequencies were reduced using standard procedures in the MIRIAD
package. The flux-density scale was set using observations of the primary calibrator,
PKS B1934$-$638, whose flux density was adopted to be 14.9 and 11.6 Jy
respectively, at 1378 and 2368~MHz. Antenna complex gains were initially
calibrated using data recorded during the 
frequent observations of the nearby secondary calibrator PKS~B0332$-$403. 
The brightnest component in the double radio source at the observing
frequencies is the SW hot spot.  The visibility data corresponding to 
the single pointing centered near this bright hot spot was initially imaged and
these visibilities were iteratively self-calibrated; subsequently these
self-calibration gain corrections were applied to all the pointing data.
At 1378~MHz several rounds of phase-only self calibration were
applied, followed by one round of  amplitude and phase self-calibration.  At
2368~MHz only one round of phase-only self-calibration was applied,
further rounds did not noticeably alter the image. Images of
all individual pointings at 1378 and 2368~MHz were deconvolved using
the Clark clean algorithm. The deconvolved images were smoothed to a common
resolution and then combined in a linear mosaic process using the MIRIAD routine
LINMOS. The primary beam attenuation over the entire mosaic image was 
corrected during the computation of the linear mosaic.
\par
The final image output by LINMOS was made as a weighted merge
of individual pointing images, which had been individually deconvolved; 
therefore, the resulting mosaic image is not
a true mosaic but what is usually referred to as a `cut-and-paste' mosaic. 
A true joint mosaicing algorithm could potentially image extended structures on
scales that are larger than the primary beam of individual antennas 
and comparable to the mosaic area. However, such joint-mosaicing is dynamic range
limited owing to uncertainties in the primary beam at low levels well off the axis:
this limitation is relevant while imaging large fields containing bright emission
peaks, as is the case here for \rg since the source has a bright hot spot.
We have not adopted a true joint mosaicing approach for this reason.
Nevertheless, we note that the largest angular scales expected to be 
reliably reproduced by the cut-and-paste mosaic are adequate  
because the angular scales of the largest emission
structures are expected to be less than the primary beam FWHM. 

\subsection{1378~MHz radio continuum}
\label{s:22cm}

\begin{figure*}
\includegraphics[width=\textwidth, angle=0]{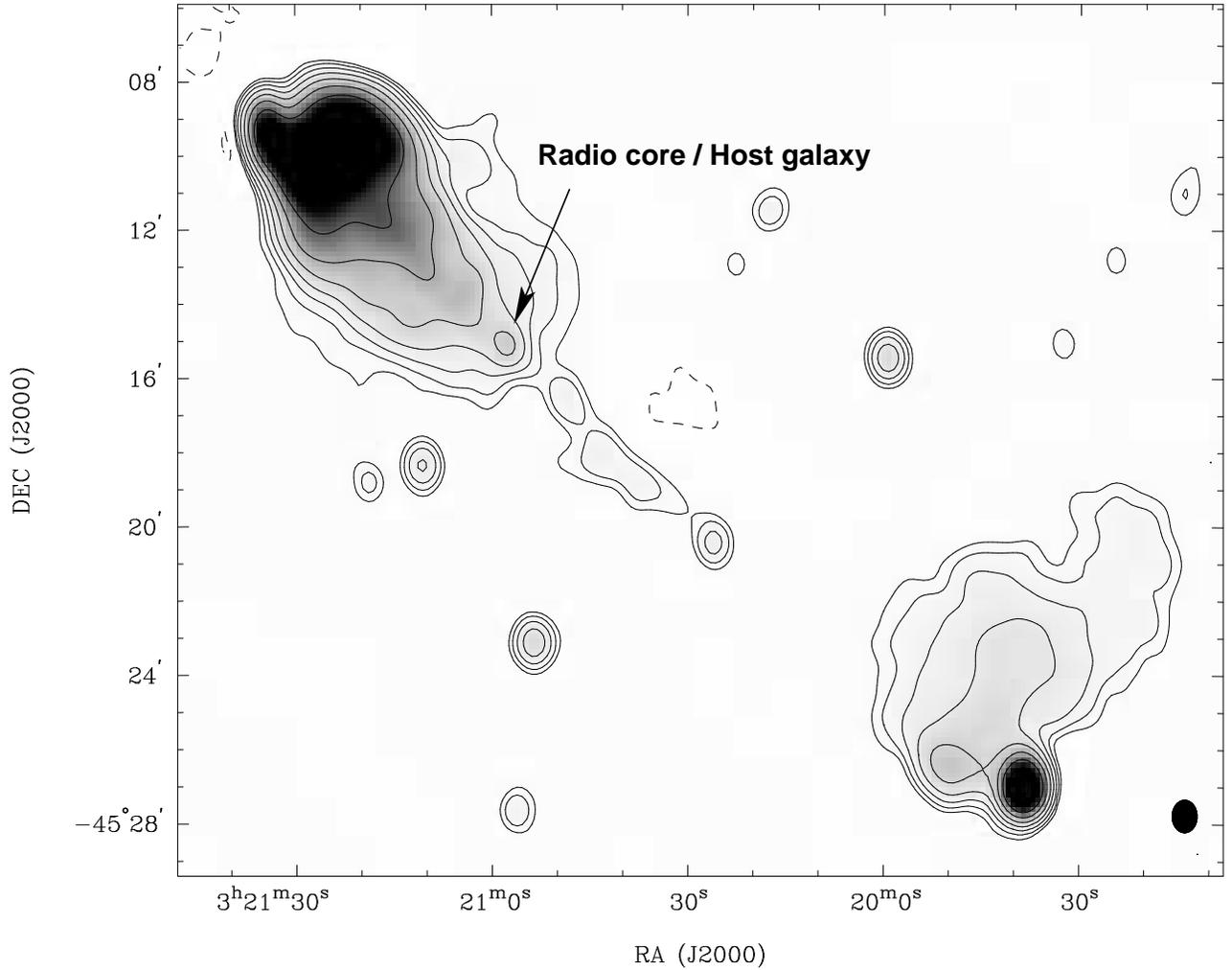} 
\caption
{\label{f:20cm} New ATCA 1378-MHz mosaic image of \rg 
made with a beam FWHM 52\arcsec $\times$40\arcsec at a P.A of $0\deg$.
Contours are at -1, 1, 2, 4, 8, 16, 32, 64, 128 and 256 mJy beam$^{-1}$. 
Grey scales are shown in the range 1-100 mJy beam$^{-1}$ using a linear scale. 
The rms noise in the image is 0.25~mJy beam$^{-1}$. The half-power size of the
synthesized beam is displayed in the bottom right-hand corner. The radio core, which 
coincides with the location of the host galaxy, is labeled. This image, as well as all 
others displayed herein, has been corrected for the attenuation due to the primary 
beam of each pointing.}
\end{figure*}

The mosaic image of \rg at 1378~MHz is shown in Fig.~\ref{f:20cm}. The image
was made with a beam of FWHM $52\arcsec\times
40\arcsec$ at a position angle of 0$^{\circ}$. The
rms noise in the image in the vicinity of the radio galaxy is 0.25 mJy beam$^{-1}$. The
image dynamic range, defined as the ratio of the peak brightness over the
entire image to
the rms noise, exceeds a 1000:1. The total intensity image of the giant radio source
shows an edge-brightened double radio source with 
two large radio lobes that are located to the north-east (NE) and south-west (SW)
of the radio core (whose location is marked in Fig.~\ref{f:20cm}). Extended emission is
also detected associated with  
a partial jet that extends from the core in the direction of the bright SW hot spot.
The jet, as in previous MOST and ATCA images is traced only over a distance of
7\arcmin~from the core, which is about one-third of the distance between the
core and SW hot spot. A weak radio source, which is associated with a
$b_{\rm J} = 17.4$ spiral galaxy \citep{jones89}, is located (in
projection) on the radio axis at the end of the observable SW jet. This object
has a redshift of $z=0.07$ and its location on the sky along the path of the
SW jet and towards the end of its visible length is likely a chance
coincidence: the galaxy  
is probably a few tens of Mpc beyond the radio source.   
\par
Two broad peaks of enhanced radio emission are located adjacent one
another and on either side of the radio axis at the end of the NE lobe. 
The ridge-like peak on the eastern
boundary is elongated parallel to the radio axis, extends further from the core
relative to the neighbouring peak located on the western boundary, and is also
slightly brighter than the second peak. It may be noted here that higher
resolution images of this lobe \citep{saripalli94} resolve the ridge along the
eastern boundary into a chain of hot spots, and suggest that the NE jet
currently terminates at a recessed hot spot located in-between the two broad
peaks seen in Fig.~\ref{f:20cm}. The post-hot spot
lobe material in this NE lobe appears to be distributed along the radio axis.
The radio core component, coincident with the host galaxy, appears to be enveloped by
this lobe plasma, at least in projection.  Our new ATCA mosaic image at
1378~MHz, with higher 
sensitivity to extended emission compared to previous images,
reveals low surface-brightness
and extended lobe material in the vicinity of the radio core. The boundary of
lobe plasma in parts of the NE lobe close to the core 
is not sharp: the wider spacing in the logarithmic radio
contours along the NW and SE boundaries 
indicate a relaxed state for this lobe material.  Additionally, the
distribution of this cocoon plasma
close to the core is not symmetric about the radio axis: the lobe is more
extended on the NW side.  
\par
The SW radio lobe is markedly different in its radio structure compared to the
NE lobe. The SW lobe is dominated by a bright  hot spot, which protrudes from
the end of a low surface brightness lobe. There is a large emission gap
between the SW lobe and the core.  A secondary peak of enhanced
emission, also noted in previous MOST and ATCA images, is observed on
the eastern boundary of this SW lobe and recessed from the bright hot spot;
this may be the site of a past hot spot or the site where the current jet bends
before terminating at the bright hot spot.
\par
The expansion and
movement of the post-hot spot plasma in this SW lobe is extremely asymmetric
about the source axis: the flow appears directed to the
NW and perpendicular to the radio axis. 
Our new 1378-MHz ATCA mosaic image clearly reveals an extension to the
low surface brightness cocoon material, perpendicular to the radio axis and in the NW
direction. The boundary of the SW lobe appears to be
relatively sharply bounded along the SE. Towards the NW, although the boundaries of the
extended emission appear to be defined, the lobe surface brightness
fades away more gradually into the plume-like extension.
\par
The $26\arcmin$ angular size corresponds to a projected linear size of 1.9~Mpc.
We measure the total flux density at 1378~MHz to be 3.86~Jy for the giant radio galaxy,
which is in close agreement with the 1472-MHz flux density measurement in
\citet{saripalli94}. The implied radio power is $4\times10^{25}$ W Hz$^{-1}$. 
Considering the absolute R-band magnitude of the host galaxy, $M_{R}  =
-23.7$ \citep{saripalli94}, \rg is placed slightly below the FR I/FR II
dividing line and in a region populated by FR I objects in the distribution
in \citet{ledlow96}.  
However, it is relatively common for giant sources, with FR II morphologies,
to have radio powers near the FR I/FR II transition \citep{ishwara99}. Also, \rg is 
a high excitation radio galaxy --- an optical spectrum of the host galaxy
shows strong narrow emission lines \citep{Bryant00}. The host
galaxies of FR-II radio sources tend to have high excitation lines in their
optical  spectra \citep{hardcastle07}; in this respect the emission line properties
of \rg are consistent with its morphological classification as an FR-II radio galaxy.

\subsection{2368~MHz radio continuum}
\label{s:13cm}

\begin{figure*}
\includegraphics[width=\textwidth, angle=0]{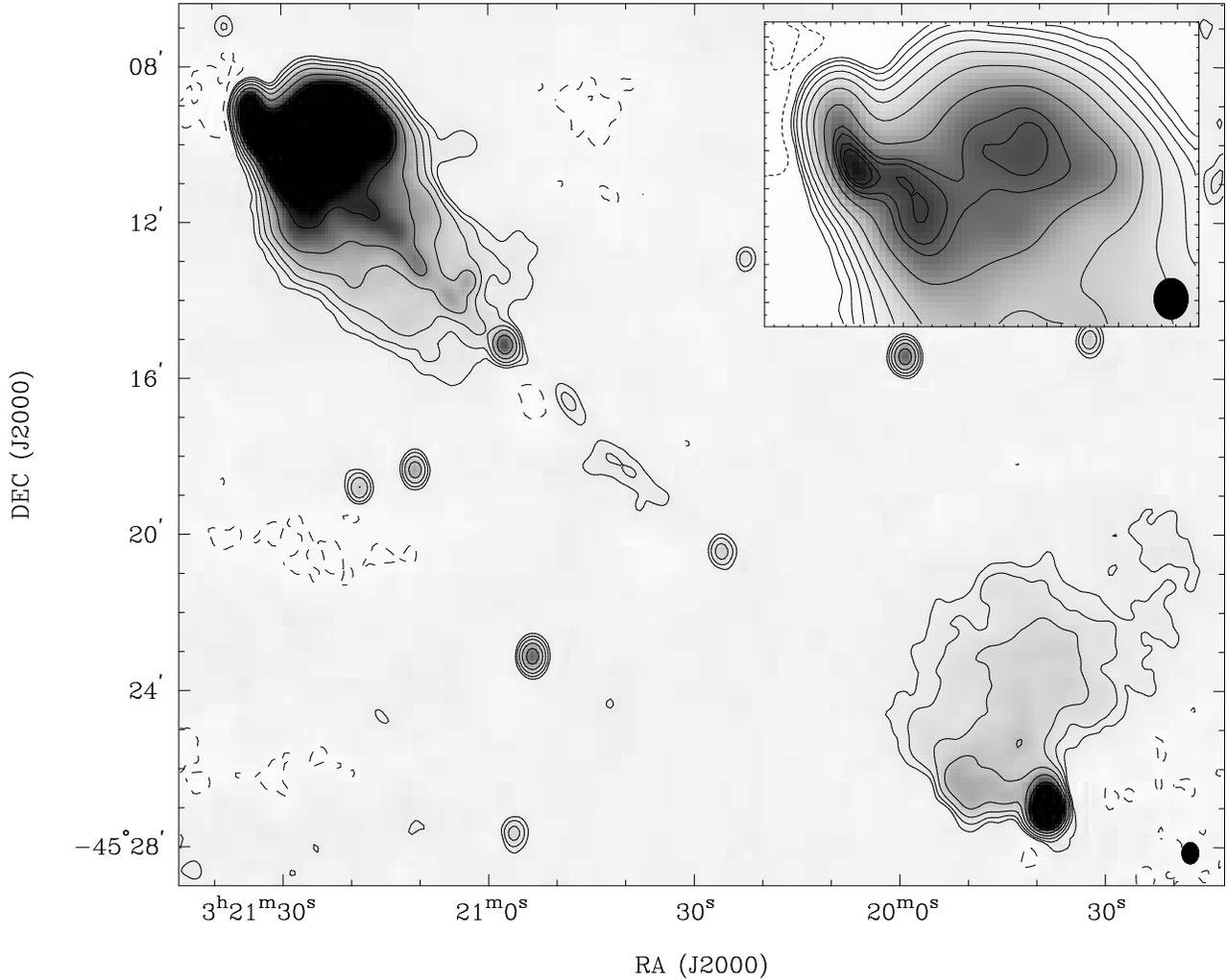} 
\caption
{\label{f:13cm} New ATCA 2368-MHz mosaic image of
0319$-$454 made with a beam FWHM $32\arcsec \times 25\arcsec$ at a P.A of 
0$^{\circ}$.
Contours are at -1, -0.5, 0.5, 1, 2, 4, 8, 16, 32, 48, 64 and 128 mJy beam$^{-1}$. 
Grey scales are shown in the range $-1$-20 mJy beam$^{-1}$ using a linear scale. 
The rms noise in the image is 0.15 mJy beam$^{-1}$. The half-power size of the
synthesized beam is displayed in the bottom right-hand corner. The inset shows the hot spot complex at the end of the NE lobe. Contours are at -1, -0.5, 0.5, 1, 2, 4, 8, 16, 32, 48, 53, 60, 64 and 128 mJy beam$^{-1}$. Grey scales are shown in the range $-1$-80 mJy beam$^{-1}$ using a linear scale. }
\end{figure*}

Our new 2368~MHz ATCA mosaic image of \rg is shown in
Fig.~\ref{f:13cm}. The image was made with a beam FWHM of $32\arcsec\times
25\arcsec$ at a position angle of 0$^{\circ}$. The rms noise on the image, in
the vicinity of the giant radio galaxy,  is 0.15 mJy beam$^{-1}$ and the image
dynamic range exceeds 1000:1. We measure the total 
flux density at 2368~MHz to be 2.46~Jy.  This is notably higher than the
corresponding flux 
density measurement of 2.04~Jy in \citet{saripalli94} and implying that
the earlier 13-cm band ATCA image may have 
missed a significant fraction of the radio emission.   
\par
We find good agreement between the previous and new ATCA images made in the
13-cm band. 
Notably, both show the same multiple hot spot complex at the end of
the NE lobe. Our lower resolution image reveals 4 peaks at the lobe
end (see the inset in Fig.~\ref{f:13cm}). The brightest and most compact  of
these is located at the very tip of the lobe, and in projection, is seen to
protrude from the lobe end. Additionally, as would be expected from the
increased total flux density in the new images, our 13-cm
band image presented here recovers some of the extended low-surface-brightness lobe
emission in the vicinity of the radio core, which was not detected in the
previous 13-cm ATCA image. 
\par
The extended structure of the giant radio galaxy as observed in the 13-cm band
image compares well with that in the 20-cm band image; however, the lowest
surface brightness features seen in the 20-cm band image are barely imaged at
13~cm. For example, a slight flare
in the lowest contour along the NW edge of the NE lobe is observed
in our 13-cm band image: the asymmetry in the NE lobe in the vicinity of the
core is not as prominent in the 13-cm band compared to the 20-cm band.  
The 13-cm band image also hints at the extension of
low-surface-brightness material in the SW lobe towards NW: once again,
although this faint 
material is reliably detected in the 20-cm band image, it is relatively
fainter at 13-cm. 

\subsection{Polarization and rotation measure}
\label{s:polrm}

Polarization images at 1378 and 2368~MHz were constructed from CLEANed Stokes
I, Q, and U images made with a beam of FWHM $52\arcsec\times40\arcsec$ at a P.A of
0$^{\circ}$. The polarization position angle images at the two frequencies were used to
compute the distribution of rotation measure (RM) over the giant radio source.
Since the difference in the orientations of the 
polarization vectors at 1378 and 2368~MHz is observed to be small, we have
assumed that our RM estimates do not suffer from $n\pi$ ambiguities. 
The computed magnitudes of RM are small over the entire 
source: mean RM is  0 rad~m$^{-2}$ and the
1-$\sigma$ scatter is less than 7 rad m$^{-2}$.  
It may also be noted here that our low RM values are consistent with
those expected for this line of sight through the Galaxy in the analysis of 
\citet{simard80}. Additionally, within the errors, we do not observe any significant
gradients in the RM distribution across the source.
\par
\begin{figure*}
\centering
\includegraphics[width=9.8cm, angle=0]{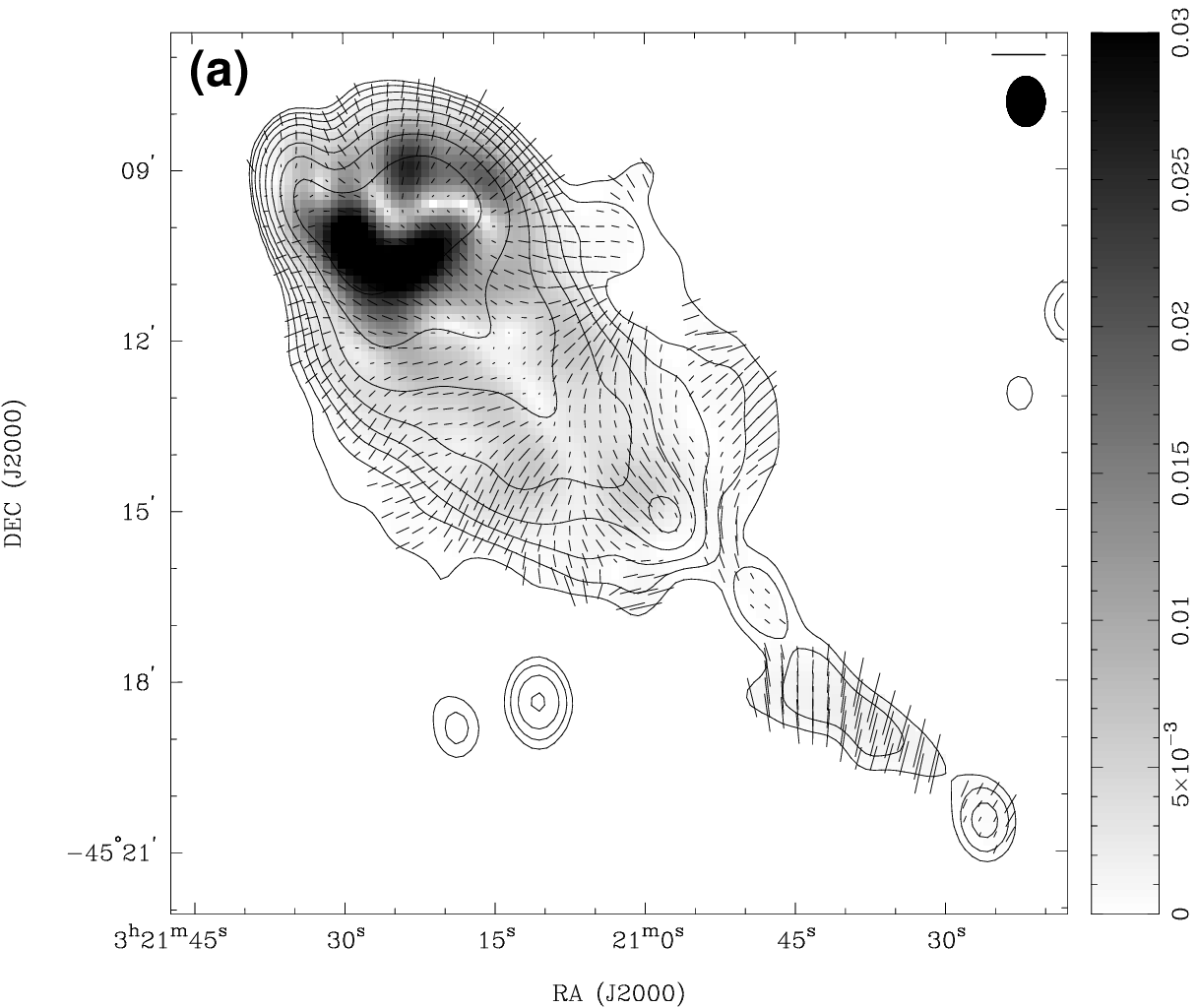} 
\includegraphics[height=9.8cm, angle=0]{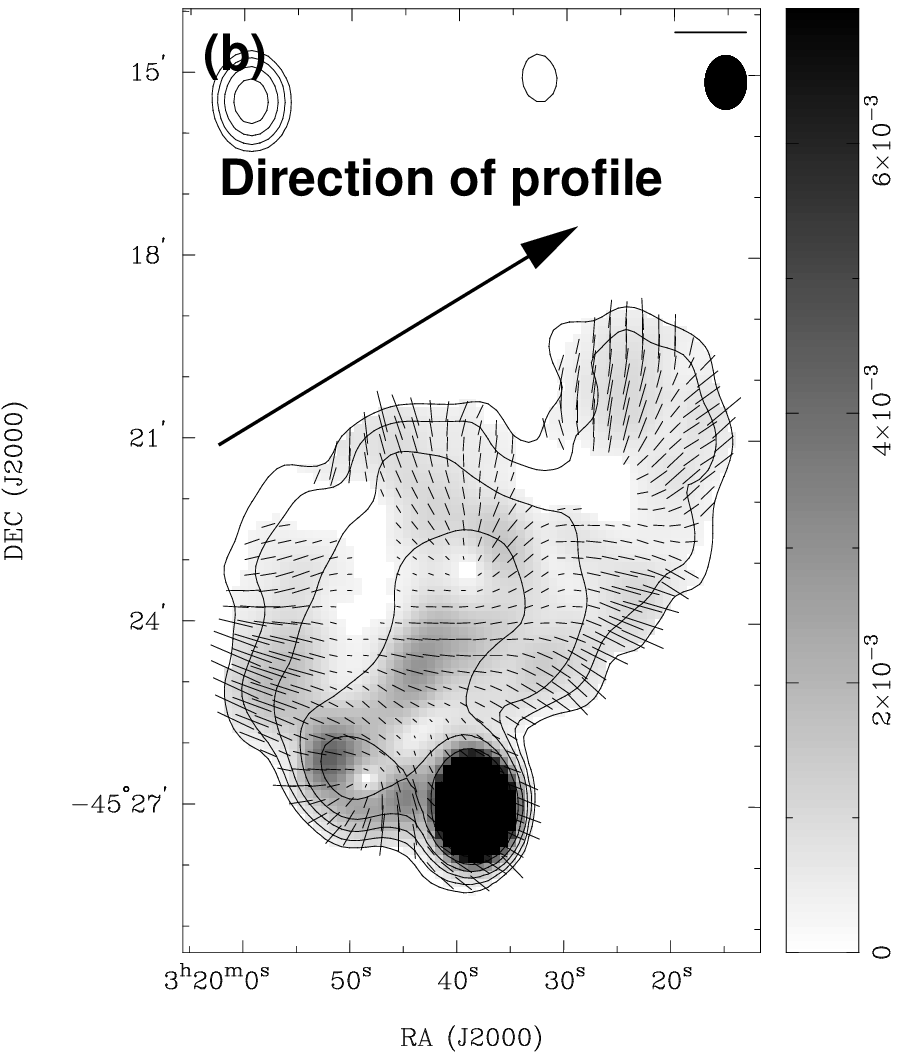} 
\caption
{\label{f:pol} Distribution of the 1378-MHz polarized intensity over the NE lobe (panel a) and
SW lobe (panel b) as observed with a beam of FWHM $52\arcsec\times40\arcsec$ at a P.A. of $0^{\circ}$. Electric field vectors are displayed with lengths proportional to the fractional polarization at 22-cm wavelength. The length of the vectors shown in the top right-hand corner corresponds to $100\%$ polarization.  Contours show the 1378-MHz total intensity  at  -1, 1, 2, 4, 8, 16, 32, 64, 128 and 256 1 mJy beam$^{-1}$. The half-power size of the synthesized beam is displayed in the top right-hand corner.
The grey-sacle is in the range 0--30 mJy beam$^{-1}$ and 0--7 mJy beam$^{-1}$ in the NE and SW lobe panels respectively. The arrow shows the direction of the mean polarization profile displayed in the following figure.}
\end{figure*}

The distribution in the 1378~MHz polarized intensity is shown in
Fig.~\ref{f:pol}. Overlaid are vectors showing the observed orientations of
the projected electric field ($E$-field) with bar lengths proportional to the
fractional polarization. The small scatter in RM over the source 
implies that the intrinsic position angles of the electric field are within 
about 10$^\circ$ of the observed angles. The polarized intensity image 
shows that there are two regions of relatively intense polarized emission at
the end of the NE lobe, which coincide with the broad peaks in total intensity.  
The fractional polarization in these regions is about $20\%$ 
at our resolution. A narrow channel of low polarized intensity and low fractional 
polarization runs between the two regions. Along this channel the polarization 
vectors sharply change in position angle; therefore, the channel is likely a 
result of beam depolarization \citep{haverkorn00}.
Another channel of low polarization is observed 
along the radio axis in the region mid-way between the radio core and the end 
of the NE lobe. To each side of this channel there are two rails of polarized emission
where the fractional polarization is about $20\%$. The projected magnetic field lines, 
which are perpendicular to the displayed $E$-vectors, follow the total intensity 
contours over most of the NE lobe excepting the central parts. The magnetic field is
circumferentially oriented along all of the boundaries at the far end 
of the NE lobe, as well as along the edges of the lobe close to the core
component, where enhancement in the fractional polarization is also
observed.  
\par
\begin{figure*}
\centering
\includegraphics[width=\textwidth, angle=0]{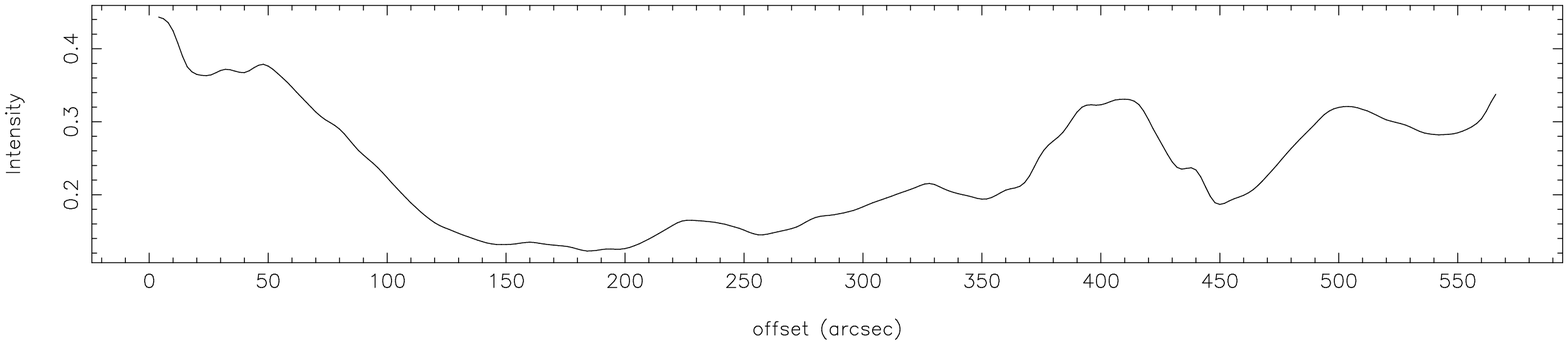} 
\caption
{\label{f:polpro} Profile of fractional polarization in the  SW lobe. The
direction of the profile is indicated in the previous figure.}
\end{figure*}

The distribution of polarized emission over the SW lobe is very similar to the
total intensity radio structure. There is a bright peak in polarized emission
at the location of the bright hot spot and there are weaker
peaks in polarized intensity recessed from the hot spot. 
Over the extended low surface brightness
regions of the SW lobe, the distribution of polarized emission is fairly
uniform and the fractional polarization is about $15\%$. The fractional
polarization is enhanced at the total intensity hot spot as well as at
the lobe boundaries. The fractional polarization values are particularly high,
and in the range 30-$50\%$, along the eastern and SW edges. 
We have collapsed our image in 
fractional polarization of the low-surface-brightness SW lobe---excluding
the hot spot---along a direction
perpendicular to the source axis. The resulting profile (Fig.~\ref{f:polpro}),
shows that the fractional
polarization is enhanced at the SE end, has a shallow minimum in the central
regions of the lobe, then increases towards the NW.
However, there is a second dip in the fractional polarization at the
base of the extension observed in this lobe towards NW and 
within the extension the fractional polarization
values increase again to about 30\% at the end.
\par
We have computed the average depolarization ratio (DR; a ratio of the
fractional polarization at 1378~MHz to that at 2368~MHz) over different regions of
the giant source.  The ratio is about  1.0 at the bright ends of the NE and SW
lobes. In the NE lobe, the depolarization ratio is about 0.9 in the central
regions of the lobe away from the hot spots and decreases to 0.75 in the vicinity of the radio
core. In the SW lobe, the depolarization ratio is $\approx1$ and fairly
constant over the low surface brightness regions, with the exception of the
NW extension where the mean depolarization ratio is around 0.6. 

\subsection{Spectral index}

We have collated radio flux density measurements for the giant radio
source from the literature and these are tabulated in
Table~\ref{t:fluxes}. The value at 408 MHz was computed from the 408-MHz all
sky survey of \citet{haslam82}; the quoted error reflects the uncertainty
in the foreground subtraction. The total spectrum for the entire source is a
power law with spectral index $\alpha\approx -0.84$ over the 
range 408 MHz -- 4850 MHz (the spectral index $\alpha$ is defined 
using the relation $S_{\nu}\propto\nu^{\alpha}$). The total spectra for 
the NE and SW lobes appear straight over the frequency range 
843--4850~MHz (where separate flux density estimates are available 
for the two lobes) with $\alpha = -0.76$ for the SW and $\alpha= -0.86$ 
for the NE lobe. We note that the 843-MHz flux density for the SW lobe 
\citep{jones89} is similar to the 1378~MHz value derived 
from our image: it is likely that the MOST measurement is missing flux density. 
At lower frequencies (80 - 160 MHz), we only have measurements 
of the flux density in the NE lobe. These are below the extrapolation
of the high frequency power-law spectrum. The measurements indicate
that the NE lobe has a flattening of the spectral index below 408 MHz; however,
it is possible that the low frequency MSH values are erroneous: \citet{mills60} 
note that the 80-MHz measurement is affected by sidelobes.
\par
\begin{figure*}
\includegraphics[width=\textwidth, angle=0]{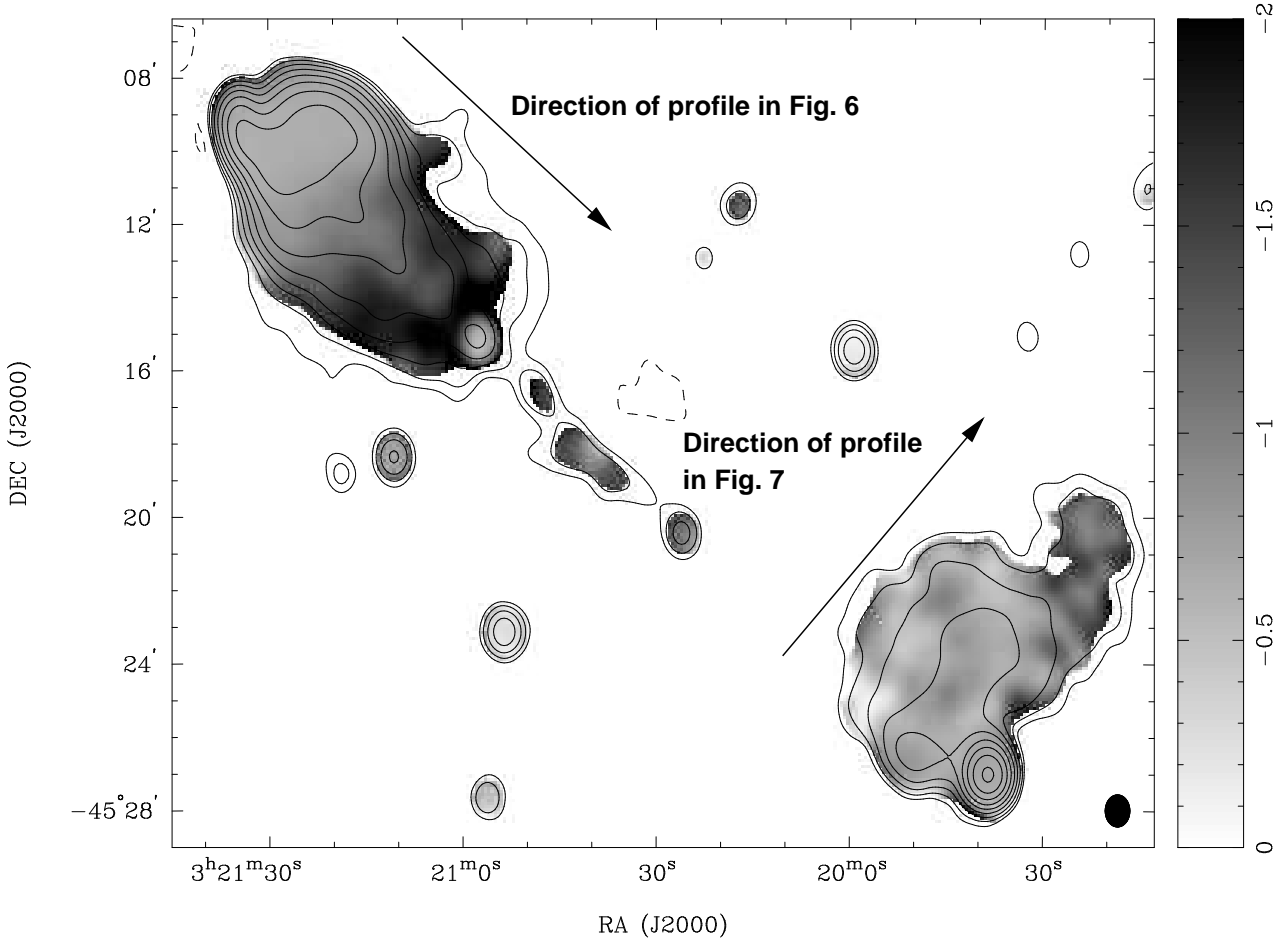} 
\caption
{\label{f:spix} Distribution of spectral index over the giant source as 
computed from images at 2368 and 1378-MHz, made with 
beams of FWHM $52\arcsec \times 40\arcsec$ at a P.A of $0\degr$.
Contours of the 1378-MHz total intensity  at -1, 1, 2, 4, 8, 16, 32, 64, 128 and 256 
 mJy beam$^{-1}$ are overlaid. The spectral index is shown 
in grey-scales in the range 0 to $-2$ mJy beam$^{-1}$ using a linear scale. 
The arrows indicate the directions of the spectral index profiles displayed in
the following figures.}
\end{figure*}

We have computed the spectral index distribution over the giant radio source
between 1378 and 2368~MHz using images with beams of FWHM
$52\arcsec\times40\arcsec$ at  P.A of 0$^{\circ}$. The resulting image is shown in
Fig.~\ref{f:spix}; the image has been blanked where the pixel values in the
individual images are less than 4 times the rms noise. 

\begin{figure*}
\centering
\includegraphics[width=\textwidth, angle=0]{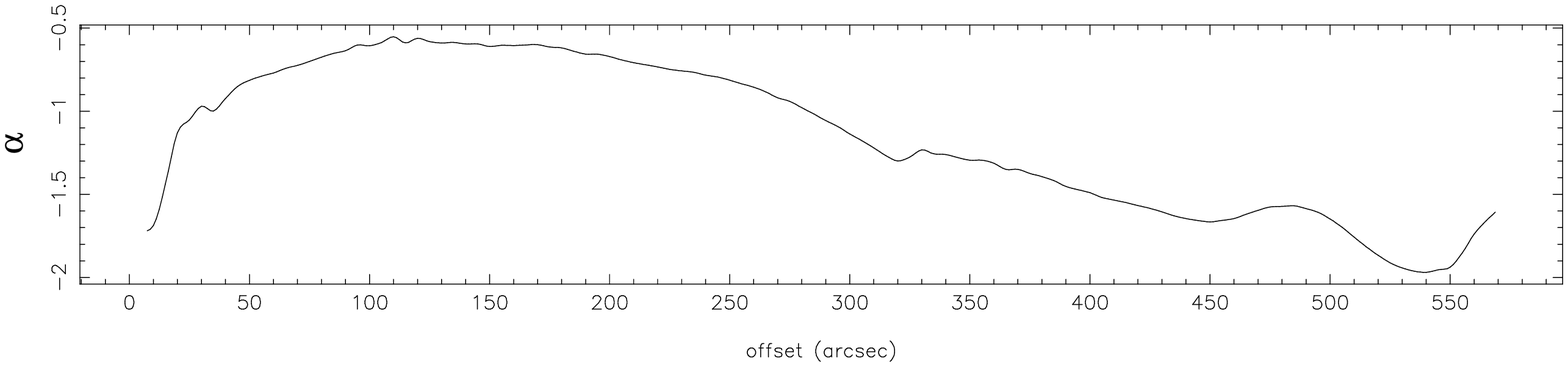} 
\caption
{\label{f:spixn} Mean profile of spectral index along the  NE lobe.}
\end{figure*}

\begin{figure*}
\centering
\includegraphics[width=\textwidth, angle=-0]{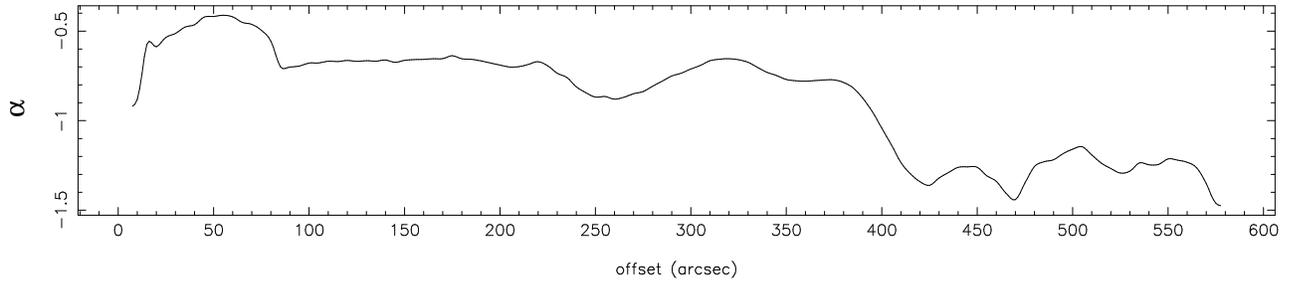}
\caption 
{\label{f:spixs} Mean profile of spectral index across the SW lobe.}
\end{figure*}

The distribution in spectral index shows a steepening of the spectral index 
along the axis and towards the core in the case of the NE lobe, and in a direction
transverse to the source axis and towards NW in the case of the SW lobe.
In both cases the steepening is along the axes of the lobes. If we assume
that older parts of the lobes have steeper spectra, as a consequence of 
spectral aging, then the observed spectral gradients suggest a backflow in the
NE lobe from the hotspots towards the core, and a transverse flow off
axis in the case of the SW lobe. We have made profiles of these observed
spectral index gradients by collapsing the spectral index distribution image
of the NE lobe perpendicular to the source axis, and that of the SW lobe along the
source axis. The profile along the NE lobe (Fig.~\ref{f:spixn}) shows that the 
spectral index smoothly steepens along the length of the lobe from 
the bright end towards the core. The profile in the SW lobe (Fig.~\ref{f:spixs})
shows an abrupt change in spectral index between the bulk of the SW lobe 
and the extension towards NW:  a step in the spectral index is observed between these
two features. The mean spectral index preceding the step is $-0.7$, 
in the extension the mean value is $-1.3$.

\section{The host galaxy and its environment}
\label{s:host}

The host galaxy, ESO 248-G10, is a luminous giant elliptical galaxy with a
warped central dust lane \citep{saripalli94}. Images of the parent galaxy at
optical and near-infrared wavelengths are consistent with a model in
which the host elliptical is triaxial with the radio axis along the minor axis
\citep{Bryant00}. 
Based on the orientation of the dust-lane, Bryant et al. argue that the
radio jet axis
makes an angle of 65$^{\circ}$ to the line of sight: the jet to the NE is pointed
away from us where as the SW jet is towards us. 

\par
In the following sections,  we present our 2dF spectroscopy of objects in the
neighbourhood of the host and use these, along with archival 6dF data, 
to infer the distribution of galaxies and
large-scale structure in the environment of the giant radio source.  

\subsection{2$^{\circ}$ field multi-fibre spectroscopy}
 
We obtained optical spectra of objects in the vicinity of \rg using the
AAOmega instrument on the Anglo-Australian Telescope (AAT), and its
predecessor, the Two degree Field (2dF) instrument. The 2dF facility is a
multi-object spectrograph, located at the prime focus of the AAT, capable of
measuring $\approx400$ simultaneous spectra within a  
2$^{\circ}$-diameter field in a single observation \citep{lewis02}. 
The AAOmega instrument, which is the successor to 2dF, is a dual-beam (blue and
red arm) bench-mounted spectrograph that uses the 2dF fibre positioner at the
prime focus. It provides greater throughput, stability and resolution than the
decomissioned 2dF spectrographs.  
\par
Our targets for the multi-fibre spectroscopy were galaxies with b$_{\rm J}$
magnitudes in the range of 15.0--19.5 selected from the SuperCOSMOS catalogue
within a 2-degree field around the host galaxy. The selection was
restricted to those objects identified to be galaxies (and not stars) in the
catalogue. A total of 1033 candidate galaxies were identified in the chosen
magnitude range. However, approximately 200 objects in the resulting target
list were previously observed in a 2dF survey of the Horologium-Reticulum
supercluster, which fortuitously covered a space-volume including \rg
\citep{klam04}. We retained about 20 of these objects in our target catalogue
as a consistency check, and omitted reobserving the others. 
\par
Two separate field configurations were prepared for the robotic fibre
positioner using the CONFIGURE routine that is part of the 2dF user software. 
For this allocation exercise, we gave priority
to objects with b$_J$ magnitudes close to that of the host
galaxy. Specifically,  objects with $15.0 < b_{\rm J} \le 16.5$  were assigned the
highest priority of 9. Decreasing priorities were given to objects in  
progressively fainter magnitude ranges: $16.5 < b_{\rm J} \le 17.5$ (priority 8),
$17.5 < b_{\rm J} \le 18.5$ (priority 7), and $18.5 < b_{\rm J} \le 19.5$ (priority 6).  The
CONFIGURE routine was also used to allocate approximately 20 fibres to blank
sky and 4 guide fibres to stars. Locations of these objects were confirmed by
examining SuperCOSMOS digitized sky survey images. 
\par
One of our two 2dF field allocations was observed on the AAT during service
time on 2003 November 26. Three exposures of 30~min each were obtained on the first
field, using the 300B grating centred at 5800A. Observations covered a
wavelength range of 3600--8000\AA~with a spectral resolution of 9\AA~FWHM. 
Unfortunately, our second allocation could not
be observed on the same night, although it had been scheduled, 
due to a field rotation problem on the
second field plate. The
second 2dF field configuration was observed with AAOmega on 2006 January
20 as part of the AAOmega Science Verification program. Four separate 20~min 
exposures were taken with the 580V and 385R volume-phase holographic
gratings in the blue and red arms respectively. These yielded spectra spanning
a broader wavelength range of  3700--8800\AA~with a resolution of $R\sim1300$  
in separate (but overlapping) blue and red halves. 
\par
2dF data were calibrated using the standard pipeline reduction package 2dFDR;
a new upgraded version of 2dFDR, for use with
AAOmega, was used for the 2006 January observation.  
Optical redshifts were determined for each object using the RUNZ software. 
Fits to templates containing typical spectral features 
were visualized: in some cases, the result of
automated fitting was clearly incorrect, usually because of residual sky lines,
and in such cases we forced the fit 
to emission line or absorption features that we considered to be
real.  Approximately $74\%$ of observed objects in our list were assigned
a reliable galaxy redshift; $13\%$ were deemed to be stars
and the remaining $13\%$ objects were considered to have unreliable
redshift estimates.  
\par
We have supplemented our measured redshifts with those from the 6dF Galaxy
Redshift (6dFGS) survey \citep{jones04a,jones05a} and redshifts from the
previous 2dF observation of the Horologium-Reticulum super-cluster,
giving a total of 664 galaxy redshifts, 92 stellar redshifts,
and 85 unreliable redshifts. The remaining 193 objects of the original
target list of 1033 objects were not allocated to a fibre in either of the
2dF, AAOmega or 6dF observations. Our completeness, over the 2$^{\circ}$
field is $73\%$ for $b_{\rm J}<19.5$.

\subsection{The galaxy distribution in the vicinity of the radio source}
\label{s:local}

\begin{figure*}
\includegraphics[angle=0, width= \textwidth]{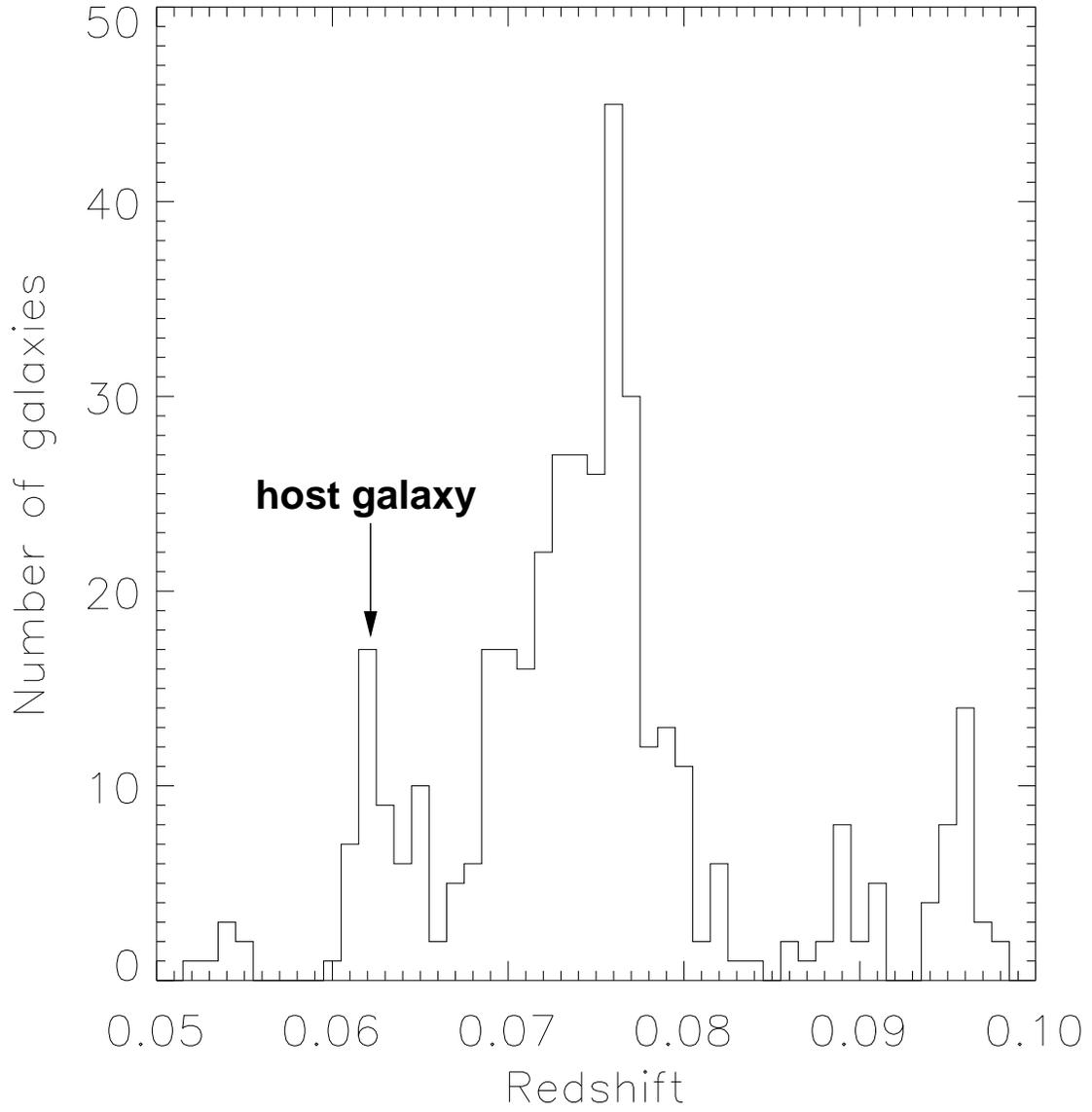}
\caption
{\label{f:zdist} Redshift distribution of galaxies within the 2-degree field in the range 
$z=$ 0.05 - 0.10. The host redshift is $z=0.0622$. } 
\end{figure*}

\begin{figure*}
\includegraphics[angle=0, width= \textwidth]{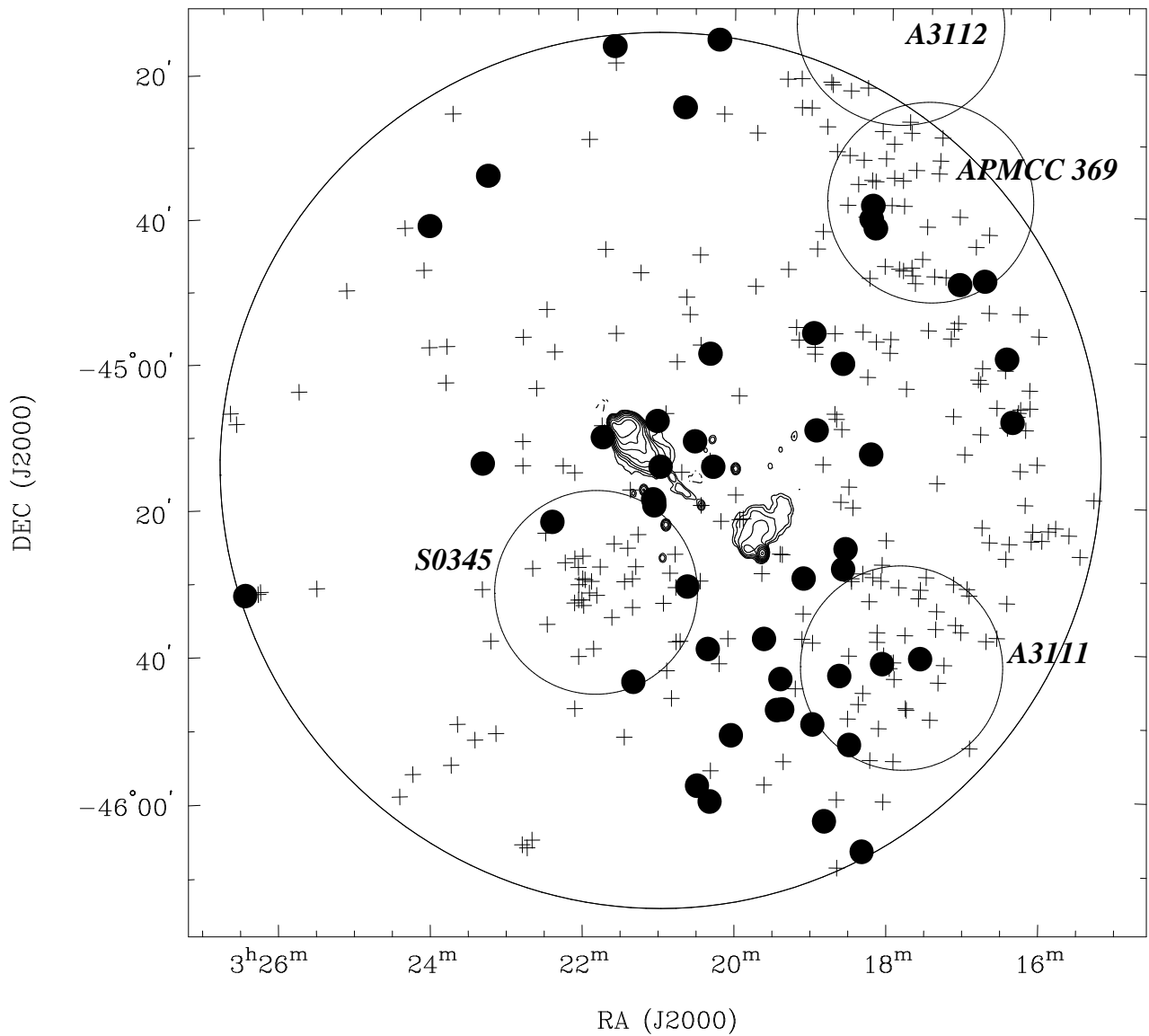}
\caption
{\label{f:skyconc} Spatial locations of galaxies in the range $z=0.060$-0.067 (filled circles) and $z=0.067$-0.082 (small crosses).  Contours show the location of the radio galaxy and are plotted at the same levels as in Fig.~\ref{f:20cm}. The large circle shows the extent of the 2 degree field. The smaller open circles show the locations of known galaxy clusters in the field within the included redshift range.} 
\end{figure*}

The redshift distribution of the galaxies with reliable redshifts 
over the range $z=0.05$ - 0.10 is shown
in Fig.~\ref{f:zdist}. The host galaxy, which
has a measured redshift of $z=0.0622$,  lies in a relatively small
concentration of galaxies that are distributed 
in the range 0.060--0.066 in redshift space. A much larger clustering of
galaxies is observed at higher redshifts and in the range
$z=0.066$ - 0.084. The sky distribution of galaxies in the two
concentrations is shown in Fig.~\ref{f:skyconc}. 
The contours in the plot show the location of the
radio galaxy and the open circles denote the positions of known galaxy
clusters within the field: these are S0345, A3111, A3112 and APMCC~369 and their
redshifts are $z=0.071$, 0.078, 0.075 and 0.075 respectively  \citep{fleenor06}.
Filled circles
mark the locations of galaxies that are within $\pm0.004$ of the host
redshift (all the galaxies in the redshift space concentration containing the
host); small crosses mark the sky positions of galaxies in the 
larger concentration in the higher redshift range. 
In both redshift ranges there appears
to be a deficit of galaxies to the NE side of the radio source.
\par
\begin{figure*}
\includegraphics[angle=0, width= 0.65\textwidth]
{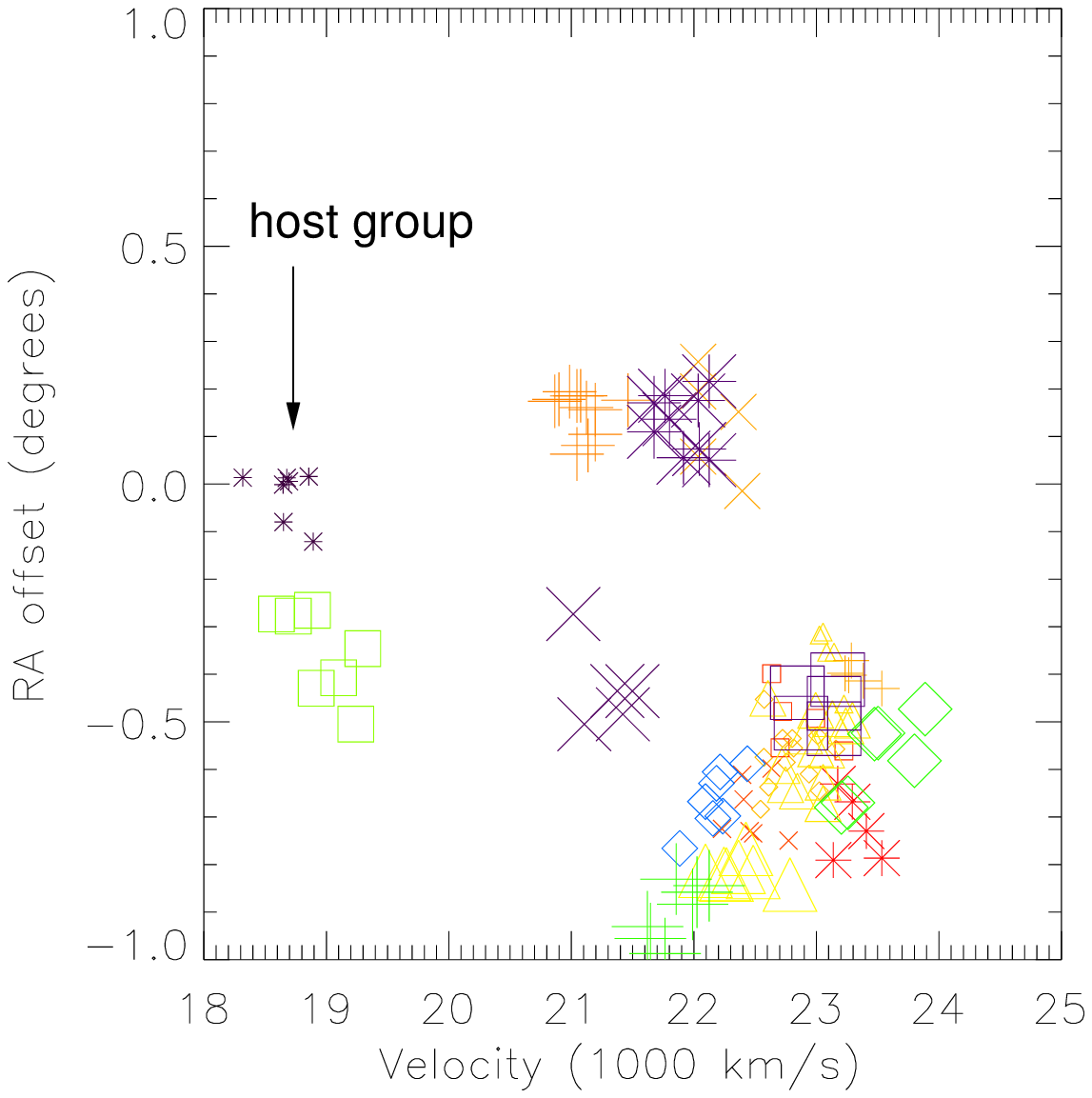}
\includegraphics[angle=0, width= 0.65\textwidth]
{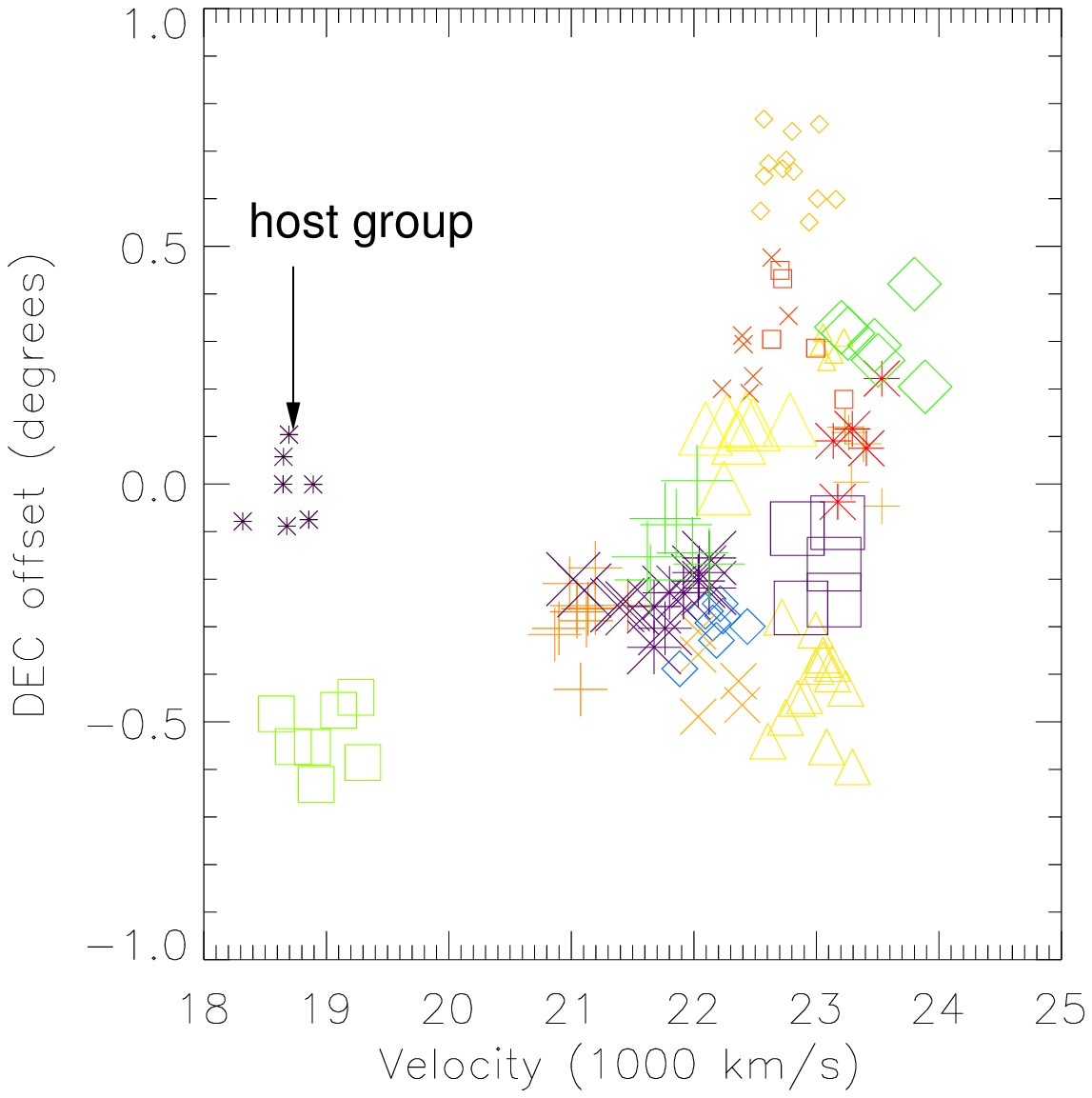}
\caption
{Locations of galaxy groups with 5 or more members 
in R.A-velocity space (top panel) and declination-velocity
space (bottom panel). R.A and declination are plotted relative
to the host galaxy. Each group has been assigned a 
different symbol and/or colour.}
\label{f:groups}
\end{figure*}

In order to identify galaxy concentrations, including any that 
might be associated with the host, ESO
248-G10, we applied the \citet{huchra82} friends-of-friends group finding
algorithm to our galaxy redshift dataset. In this technique, association is
determined by the projected spatial and velocity separations between
individual galaxies. In particular, a galaxy is a friend of another galaxy if
it is within a limiting projected distance $D_{\rm L}$ of that galaxy, and if
the magnitude of the velocity difference between the two galaxies is less than
$V_{\rm L}$. Large values of $D_{\rm L}$ and $V_{\rm L}$ result in most of the
galaxies included into large associations comprising multiple groups strung out
in velocity space.  Very small values result in the identification of tight
groups and sub-groups with most galaxies excluded from the associations.  We
explored a range of parameters, systematically reducing the parameter space
from $D_{\rm L}=4$ Mpc and $V_{\rm L}=1000$ km s$^{-1}$ until the identified groups
appeared as compact separate features in plots of R.A. versus declination,
R.A. versus velocity, and declination versus velocity. 
The chosen values,  $D_{\rm L}=1$ Mpc and $V_{\rm L} = 400$ km s$^{-1}$, resulted in the identification of the galaxy concentrations shown in Fig.~\ref{f:groups}. 
For clarity, isolated galaxies and groups with less than 5 members are
not plotted in this figure.
The host of the radio source is in a concentration with seven identified members. 
Two of the members appear close together in SuperCOSMOS
optical images of the field, and there is a third galaxy, of unknown
redshift, also located within 10 arcseconds of one of these on the R-band
image. A second concentration, also consisting of seven identified members in
our redshift database, is located about $40\arcmin$ to the SW
of the concentration around the host and at a similar redshift. 
All other concentrations identified by
our friends-of-friends algorithm are offset from
these two in velocity space by $>2000$ km s$^{-1}$. It is interesting
to note that all the identified galaxy concentrations are located
towards the SW side of the 2$^{\circ}$ field, only isolated galaxies occupy
the sky region to the NE. 
\par
The host galaxy, with apparent magnitude $b_{\rm J}=16.10$, appears
to lie close to the centre of 
the associated concentration on the sky. The host galaxy is the brightest
member of this concentration, the second brightest galaxy  
has a magnitude $b_{\rm J}=16.25$ and the other members are fainter
than $b_{\rm J}=17.2$. The concentration is distributed over a projected
linear extent 
of about 1~Mpc on the sky, indicating that this might be a loose group.
We measure a velocity dispersion of 196 km s$^{-1}$ for the galaxies 
identified with this concentration
and also note that the host lies close to the centre of the
distribution in velocity space.
The concentration to the SW is distributed over a larger sky area and 
has a larger velocity dispersion, $\sigma =
272$ km s$^{-1}$, and has a mean velocity that is 290 km s$^{-1}$ higher than
that of the concentration around the host. We have examined the ROSAT All Sky
Survey archival 
data at the location of the 2$^{\circ}$ field and do not find any X-ray
emission associated with the host grouping or with the neighbouring 
grouping to the SW.
However, there is low surface brightness extended emission
coincident with the position of the distant $z=0.071$ Abell cluster S0345.
\par
\begin{figure*}
\includegraphics[angle=0, width= 0.5\textwidth]{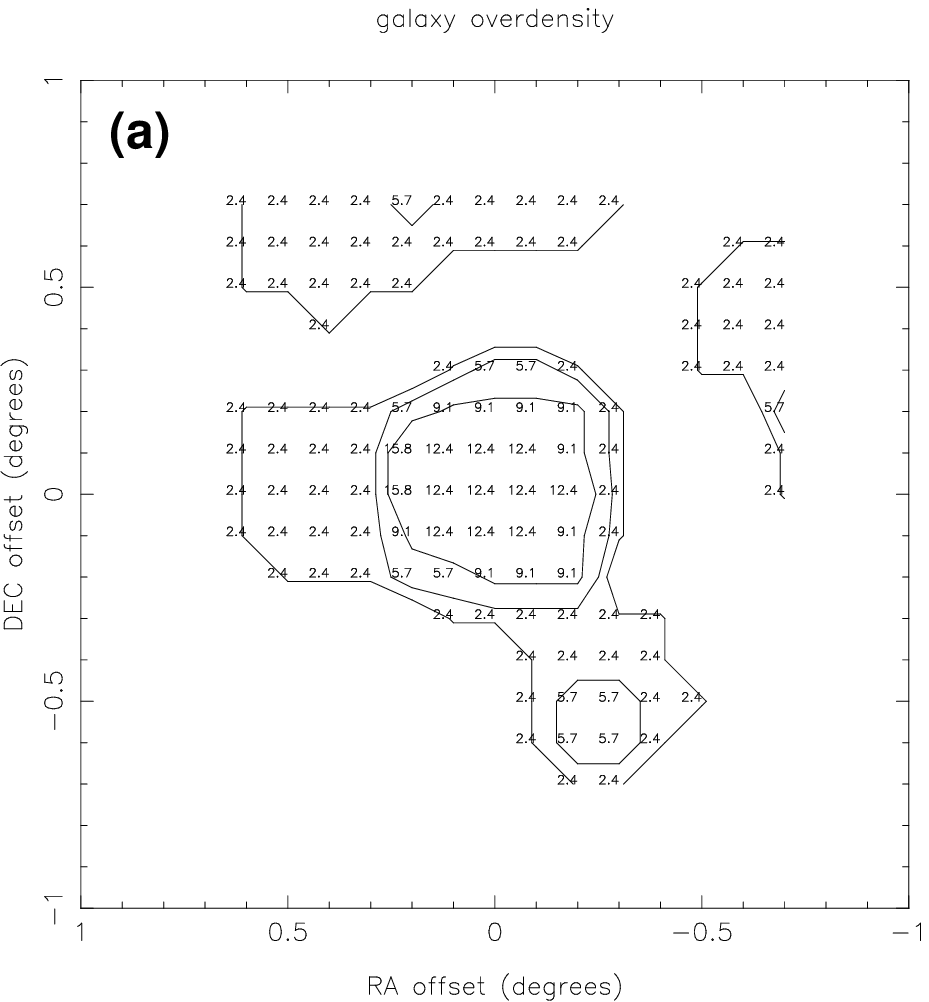}
\includegraphics[angle=0, width= 0.5\textwidth]{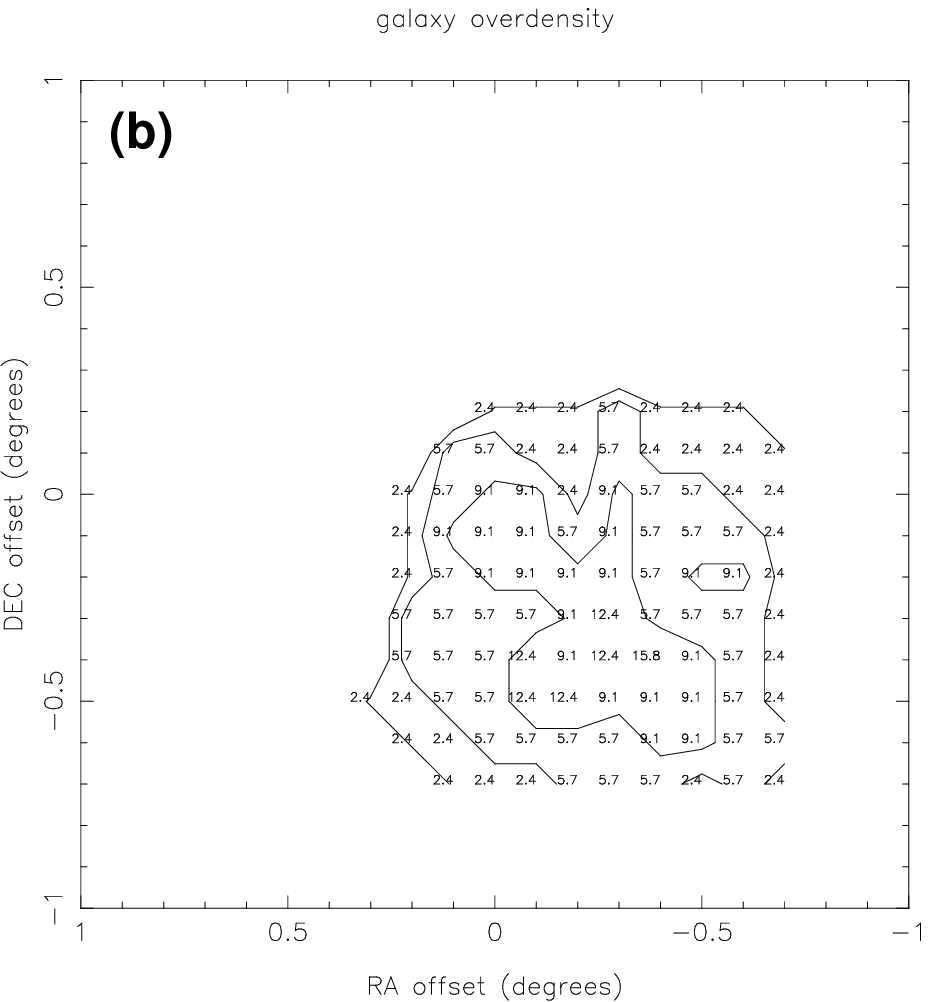}
\caption
{Fractional over-densities in the vicinity of the radio source.
Contour levels are shown at $\Delta n/\bar{n}$= 2, 4, and 8.
Panel (a) is in a sky plane at the location of the host galaxy, whereas panel (b)  
is off-set in line of sight distance by 3.8 Mpc. The fractional over-densities displayed in
panel (b) are behind the host galaxy. 
R.A and declinations are off-set relative to the host galaxy position.}
\label{f:lods}
\end{figure*}

We have computed the 3-D spatial distribution of galaxy number density
in the vicinity of the giant radio galaxy. In this anaylsis we included all
galaxies within a box of dimensions 6 Mpc $\times$ 6 Mpc $\times$ 25 Mpc,  
which included the two concentrations of galaxies
in the redshift range 0.060--0.066. 
Galaxies within $\pm0.7^{\circ}$ in R.A. and declination and
with redshift offsets within $-0.0022<\Delta z < 0.0038$ 
of the host galaxy were included in the box. We neglected
peculiar velocities and assigned distances along the line of sight based on
galaxy redshifts. In order to compute the galaxy over-density, we smoothed the
data using 
a top-hat function with smoothing radius 1.25 Mpc. At this smoothing scale, the
mean number density of galaxies is 0.3 and the rms number density is 2.2.
Figure~\ref{f:lods} shows the fractional over-densities ($\Delta n/\bar{n}$)
in a sky plane at the redshift of the host. The fractional over-density at the
location of the 
host is 12.4. The peak fractional over-density within the cube is 15.8 and
is at an offset position, with respect to the host
galaxy, of $-$1.7 in R.A., $-$2.5 in declination and $+$3.8 Mpc
along the line of sight: this position is within the location of the
galaxy concentration to the SW. Fractional over-density values in the 
sky plane containing  
the peak is also shown as a separate panel in Fig.~\ref{f:lods}. 
All pixel locations with $\Delta n/\bar{n}$ exceeding 6
within the cube are either associated with the host group or with the SW group.

\par
The jets in \rg make an angle of 65$^{\circ}$ to the line of sight with the 
NE jet pointed away from us. We have examined the fractional over-density
values in the box, with smoothing radius 1.25~Mpc, along 
the inferred axis of the giant radio source.  $\Delta n/\bar{n}$ is fairly 
constant over the region of the NE lobe, and the value of 
$\Delta n/\bar{n}$ somewhat exceeds 10. 
The SW lobe is located outside the extent of the concentration 
with which the host
galaxy is associated, and obviously lies in a region of significantly lower
galaxy number density: the galaxy overdensity within the cube decreases
along the path length of the SW jet and has values as low as $\Delta
n/\bar{n}=2$ in the vicinity of the SW hot spot.  If the galaxies are a 
tracer of gas, the local galaxy distribution
constitutes evidence for an asymmetry in the gas density on the two sides of
the host; additionally, the distribution is evidence for a density gradient  
in the ambient medium of the radio galaxy, between the radio core and the 
SW hot spot.
\par
We have also computed the gravitational field in the vicinity of the
radio galaxy following the method described by \citet{subrahmanyan08}.
Again, peculiar galaxy velocities have been ignored in this computation and 
distances to galaxies along the line of sight are based on redshift values only. 
First, we estimated the total mass in the cube as the product of the box 
volume, $V$, and mean matter
density in the Universe, $\rho_{\rm m}$. Assuming that the galaxies trace this
matter, an average mass, $M_{\rm g} = (V\times\rho_{\rm m})/ N_{\rm g}$ was
assigned to each galaxy,  where $N_{\rm g} = 37$ is the total number of
galaxies in the cube. This resulted in a mass of $M_{\rm g} = 2.0\times
10^{11} \msolar$ associated with each galaxy. The acceleration vector at
any location within the cube was then computed by summing the contribution 
from each $M_{\rm g}$ point mass. Nearby galaxies were excluded from the
sum, so that the computed gravitational field is that due to surrounding
large-scale galaxy structures, and not individual objects.
 Galaxies outside of the cube were also ignored.
The resultant acceleration vector at the location of the host
has components: $g_{\rm RA} = -1.2\times10^{-13}$ m s$^{-2}$, $g_{\rm DEC} =
4.7\times10^{-14}$ m s$^{-2}$, and $g_{\rm z} = 1.4\times10^{-13}$ m s$^{-2}$. 
The magnitude of this acceleration vector is $g_{\rm s} = 1.9\times10^{-13}$ m
s$^{-2}$ and the component on the sky plane is directed NW 
at a P.A of $-$68$^{\circ}$.  Assuming the age of the Universe to be $\tau =
13.7$ Gyr, the acceleration corresponds to a velocity of 
$g_{\rm s} \tau = 82$ km s$^{-1}$. 
Within the extent of the group, the magnitude of the gravitational
acceleration increases with distance from the host and takes on values about a
factor of two greater than that at the location of the host galaxy;
 outside the group the
acceleration declines with distance from the group. Within a radius of about
1~Mpc of the host galaxy, the acceleration appears to be dominated by the mass
associated with the group of which the host is a member.
At the location of the SW lobe the vector components are much smaller: $g_{\rm
  RA} = 
2.9\times10^{-14}$ m s$^{-2}$, $g_{\rm DEC} = 6.4\times10^{-15}$ m s$^{-2}$,
and $g_{\rm z} = 5.0 \times10^{-14}$ m s$^{-2}$; this corresponds to a
velocity of $g_{\rm s} \tau = 25$ km s$^{-1}$ and is roughly directed towards
the concentration associated with the host galaxy. At the location of the SW
hot spot and lobe, the gravity is not dominated by the second galaxy
concentration to the SW. 

\subsection{The large-scale structure in the neighbourhood of the radio source}
\label{s:largescale}
 
\begin{figure*}
\includegraphics[angle=0, width= \textwidth]{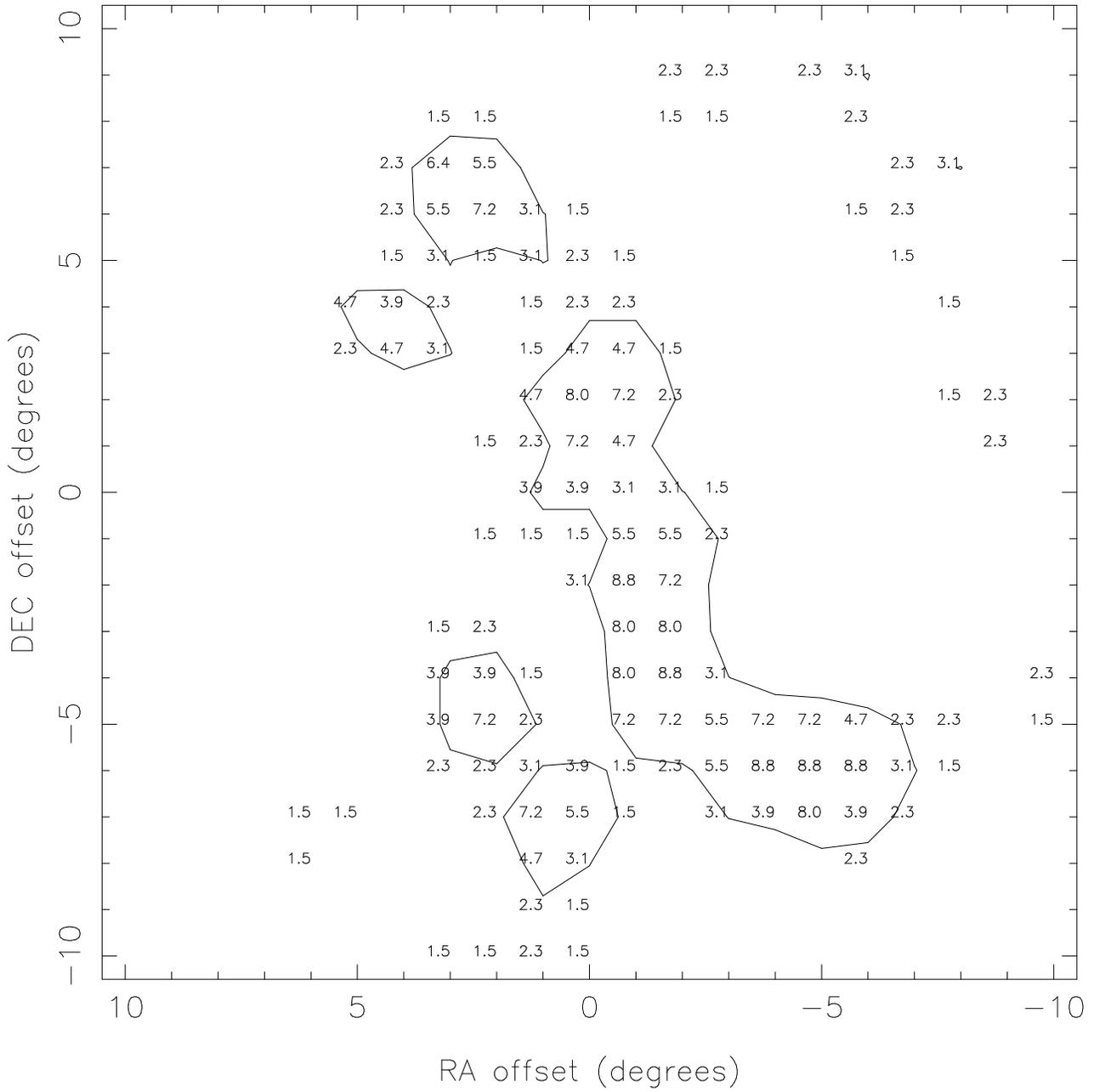}
\caption
{Numerical values of fractional galaxy over-density in a 20 deg field around the host. 
The data have been smoothed with a top-hat function of radius 6 Mpc.
R.A and declination are off-set relative to the position of the host.}
\label{f:dens6df}
\end{figure*}

We have used archival data from the 6dFGS to infer the distribution of
galaxies in the  vicinity of \rg on larger scales. We computed the 3-D spatial
distribution of galaxy number density using 6dF galaxies within a cube of
side  85 Mpc centred at the host. A total of 860 galaxies within $\pm10\deg$
in R.A. and declination and within $\pm0.01$ of the redshift of the host
galaxy were included in this analysis.  
In order to compute the galaxy over-density, we have smoothed the data using a
top-hat function with radius 6~Mpc. At this smoothing scale, the mean number
density of galaxies in the cube is 1.2 and the rms is
1.5. Figure~\ref{f:dens6df} shows the fractional over-densities ($\Delta n /
\bar{n}$) in the sky plane at the redshift of the host galaxy. 
At the location of the host, the
fractional over-density is 3.9 with this smoothing scale. 
This plot shows that the host appears to be
embedded within a large-scale galaxy filament that is oriented in a  NNE - SSW
direction. The filament extends more than 60 Mpc in our plot
and has a projected width of approximately 15 Mpc, which is the width between
points with fractional overdensity of two at this smoothing scale. Fractional 
over-density values in front of and behind the host galaxy reveal that the 
filament has an apparent depth of only 8 Mpc along the line of sight.  
However, the real physical size, in redshift space, is likely to be
larger than this because of the Kaiser effect in this filament that has a low
fractional overdensity: there may be an apparent compression
in the distribution of measured velocities due to the infall, in comoving
space, of galaxies towards
the filament centre along the line of sight \citep{kaiser87}.
In addition, the filament is observed to bend away from us in the northern 
parts of the cube and, conversely, bend and extend towards us in the southern parts.
Within the galaxy filament, the fractional over-density takes on values up to
about 9 with this smoothing scale. The host galaxy is offset from the centre 
of the filament, in redshift space, by approximately 1.5 Mpc towards us.
The total extent of the radio source---about 
1.9/sin(65$^{\circ}$)~Mpc = 2.1~Mpc---is much smaller than the width of this
large-scale  
filament of galaxies, and the entire radio source, including the galaxy concentration 
associated with the host galaxy, are embedded within and located 
in the central parts of the filament.  As projected on the sky, the radio jets
are roughly in a direction parallel to the large-scale filament; however, the
inclination of the radio axis to the line of sight makes the SW lobe somewhat
more distant from the axis of the large-scale filament.
\par
\begin{figure*}
\includegraphics[angle=0, width= \textwidth]{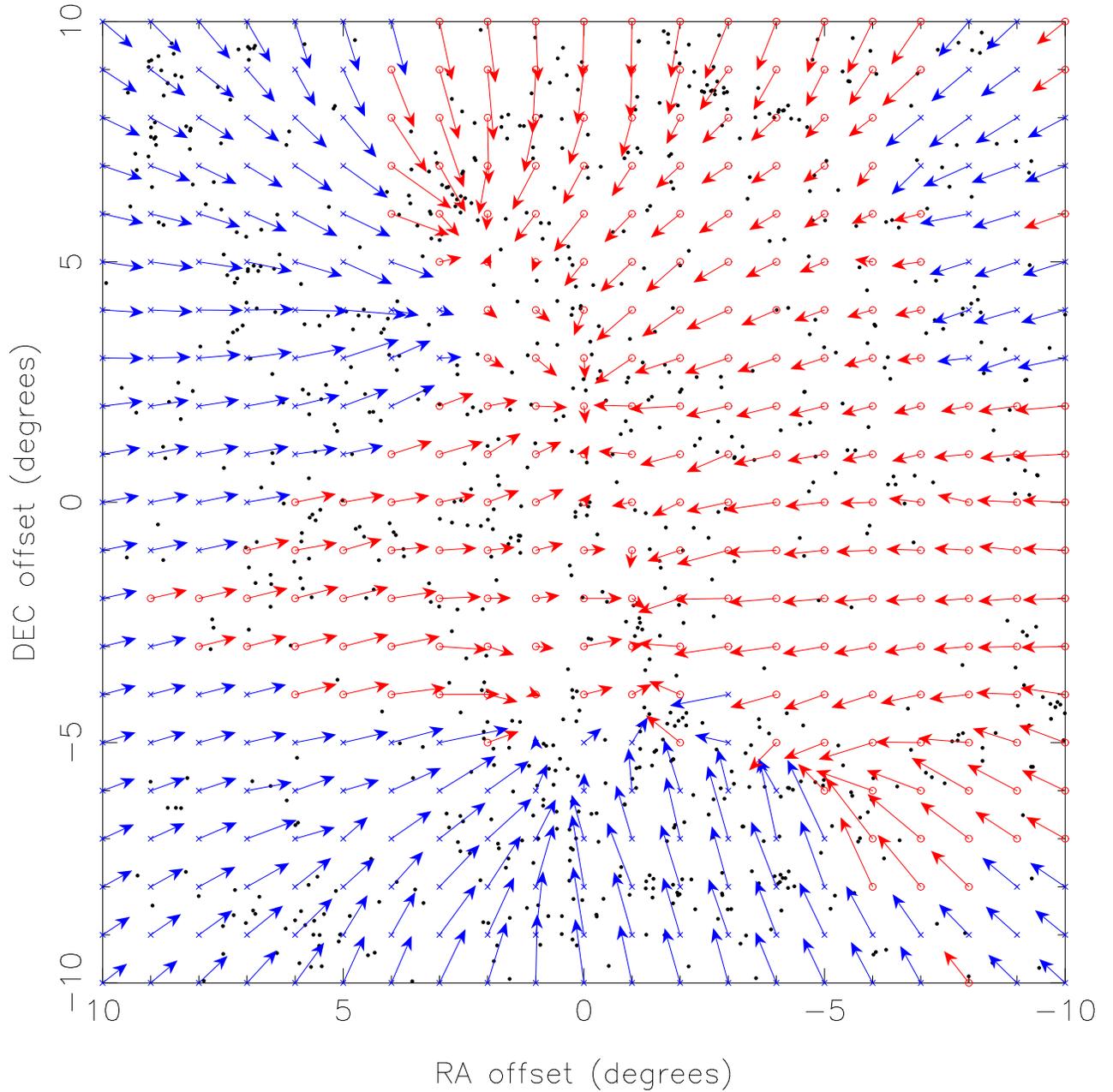}
\caption
{Gravitational acceleration vectors at the redshift of the host galaxy 
projected onto the sky. Blue vectors (with crosses) are directed out of the page, 
while red vectors (with circles) point into the page.}
\label{f:gravity6df}
\end{figure*}

Using an identical technique to that described in the previous section, we
have computed the gravitational field resulting from the large-scale structure
using the 6dF survey galaxies in the 85 Mpc side cube. The average mass
assigned to each galaxy in this cube was $M_{\rm g} = 2.6\times 10^{13}
\msolar$. In this analysis galaxies within 4.0~Mpc (almost 1 degree) of
the location at which any acceleration vector was computed were ignored. 
Therefore, the resulting acceleration vector field is that due
to the large-scale structure and not influenced by local galaxies or
concentrations. The resulting gravitational field, computed on the sky plane 
at the redshift of the host galaxy, is displayed in
Fig.~\ref{f:gravity6df}; the vector lengths represent the component on the
sky plane.  The vector components 
at the location of the host are $g_{\rm RA} = -1.9\times 10^{-13}$ m s$^{-2}$,
$g_{\rm DEC} = 2.18 \times 10^{-13}$ m s$^{-2}$, and $g_{\rm z} =  5.12 \times
10^{-13}$ m s$^{-2}$. The magnitude of the acceleration is $g_{\rm s}
=5.9\times 10^{-13}$ m s$^{-2}$, most of which is directed along the line of
sight away from us, as would be expected from the inference made above that
the host galaxy lies in front of the axis of the filament.  The corresponding
velocity is $g_{\rm s} \tau = 253$ km 
s$^{-1}$ and on the sky plane the acceleration is 
directed towards P.A. of $-41^{\circ}$. The relatively small acceleration
vector at the location of the radio source is consistent with the finding that
the host galaxy is located embedded within the filament and not in the
peripheral regions.

\section{Evolutionary history of the giant  radio galaxy \rg}
\label{s:hist}

The NE lobe in \rg is aligned along the radio axis, whereas the SW
lobe appears to be extended in a direction perpendicular to the source axis. The
contrasting shapes and orientations taken by the two lobes are indicative
of two very different flow histories. The total intensity 
structure in the NE lobe is suggestive of a model in 
which the jet axis remains fairly steady over most of its length, but has
significant jitter at the end.  The broad peak at the northern end of the lobe
was probably the 
site of past hot spots, and the chain of hot spots along the other eastern rim,
which protrude past the end of the lobe, the sites of current
hot spots. Strong and directed backflow is indicated along the source axis and
towards the core, which continues beyond the core. The backflowing cocoon
material appears to bend away from the source axis and towards NW, filling and
inflating a relatively relaxed and asymmetric bridge in the vicinity of the core.
\par
The gradual steepening of the spectral index distribution along the length
of the NE lobe suggests there are significant age gradients
along the lobe axis: the radio spectrum is steeper in the vicinity
of the radio core where the aged electron population has had more time to
lose energy via synchrotron emission. If the depolarization (see Sect.~\ref{s:polrm})
is a result of entrainment, which is higher in aged cocoon material, then the
DR distribution supports this view. 

\par 
The radio structure in the SW lobe is suggestive of a different flow history.
There is a large emission gap between the SW lobe and the radio core, and it 
appears that the movement of post-shocked plasma is to the NW and 
not along the source axis.  
Alternative models in which the plume-like feature 
to the NW trace the path of past jet termination points due to, for example,
jet precession, or models invoking movement of the host galaxy with respect to
the ambient IGM in which the jet material is deposited are unlikely because
they would be expected to manifest as symmetric or inversion symmetric lobe
distortions on the two sides.
\par
Enhancements in fractional polarization as well as an orientation of the
B-field parallel to boundaries may be interpretated as arising from a
compression of a magetized plasma, with a tangled field, at locations where
the flow terminates on ram-pressure interaction with ambient thermal plasma.
The B-field orientation is transverse to the source axis in the bright SW
hot spot, as might be expected from the interaction between the jets and
ambient gas at the termination shock. The magnetic field lines generally
follow the total intensity contours along the boundaries of the SW lobe indicating 
that the lobes are not relaxed but are compressed at the boundaries where the
expansion is ram pressure limited. Away from the hot spot and in the central
regions of the SW lobe the B-field is oriented along NW-SE indicative of a
flow of the post hot spot material towards NW.  The enhanced fractional
polarization and circumferential B-field in the regions of the SW lobe just
before the extension towards NW, manifesting in the intermediate peak in the
slice profile in Fig.~\ref{f:polpro}, suggest that the flow is discontinuous
across the SW lobe and that the extension towards NW might have a 
separate origin.
\par
A model that might be considered is one in which 
the post hot spot material inflates a lobe at the location
of the hot spot, which subsequently bends by $\approx90^{\circ}$ to be directed
along the line of sight, and then 
bends once again into the plane of the sky and towards NW to form the
extension.  In this picture, we would be observing new plasma that has
freshly been accelerated at the hot spot, and old plasma, which resides 
behind or in front of the new plasma, that in projection on the sky, appear 
to form one continuous structure oriented NW. An argument against such a model
is that the total intensity image does not show an enhancement prior to the
extension towards NW.
Also, it is difficult to explain why the lobe should suddenly bend in this
way. One possibility, is that the backflow is deflected by the thermal gaseous
halo associated with the galaxy group of which the host is a member, in
the same way that backflows are deflected by large angles when they encounter
galaxy halos \citep{kraft05}.

\par
The total radio spectrum for the SW lobe is straight over the frequency range
1378-4850 MHz. It flattens at 843 MHz, but as noted earlier, this is probably
because this observation missed some of the flux density.
Values of the two point spectral index, $\alpha^{\rm 22}_{\rm 12 cm}$, over the 
SW lobe take on values $\alpha\approx-0.7$ over the bulk of the emission 
and steepen to $\alpha\approx -1.3$ in the lobe extension to the NW. We assume 
that the observed spectral index of $-0.7$ over the relatively higher surface
brightness parts of the SW lobe represents the injection spectrum, and infer
that there is no break, over the frequency range 843-4850 MHz, in the spectrum of 
the radiating electrons that populate the head and bulk of the SW lobe. Therefore, 
any steepening in the spectrum, as a result of radiative aging of the electron 
population, must occur at frequencies $>4850$ MHz.  
The SW lobe has a minimum energy magnetic field strength $B_{\rm me} = 0.1$ nT. 
The equivalent magnetic field strength of the cosmic microwave background 
at the redshift of the source is $B_{\rm MB} = 0.45$ nT; therefore, 
particle aging is dominated by  inverse Compton losses due to the microwave
background radiation and not synchrotron radiation.
This is typically so for giant radio sources, which have large
expanded lobes and low equipartition magnetic fields \citep{ishwara99}. 

\par
Breaks in the emission spectrum due to both
synchrotron and inverse-Compton losses are expected to occur at a frequency
\begin{equation}
\nu_{\rm T} = 1.12\times10^{3} \frac{B_{\rm synch}}{(B^{2}_{\rm synch} +
  B_{\rm MB}^{2})^{2}t^{2}}  
\mbox { GHz,}  
\end{equation}
where t (in Myr) is the spectral age or time since acceleration. We estimate 
a spectral age $t<2.1\times10^{7}$ yr for the bulk of the SW lobe. The steeper two-point 
spectral index, with $\alpha^{\rm 22}_{\rm 12 cm}\approx -1.3$, 
in the extension to the SW lobe suggests an aged plasma. A plausible model
for this extension is that if the injection spectrum 
had an emission spectral index $\alpha = -0.7$, the
population might have aged over a time exceeding $>4\times10^{7}$ yr while
experiencing  continuous injection of reaccelerated electrons---resulting in a
spectral break at frequencies $< 1378$~MHz---then undergoing passive 
aging over a period of at most $3\times10^{7}$~yr.

\section{Processes driving the lobe evolution}
There are a number of physical models that can relate the observed extended
emission structure of the radio source and the ambient environment. The
environmental influence may be in the form of
the gravitational field of the large-scale distribution of matter, and the gas
associated with the large-scale structure.  We assume here that the galaxies
trace the mass as well as the gas. In the following sub-sections,
we consider some models that could potentially be relevant to the evolution of the radio
structure. We examine their relevance to this case study, 
in an attempt to shed light on the physical  
processes leading to the formation of the asymmetries in \rgns.

\subsection{Buoyant backflows}

The morphology of the SW lobe is plume-like, directed NW and moving 
away from the radio axis. We model this lobe as a backflow from the 
hot spot that is initially directed towards the core, but is deflected off the source 
axis and towards the NW because of buoyancy. The buoyant forces are a result 
of a gravitational field whose direction is transverse to the source axis. 
We have shown previously that the gravitational field on local scales
is largely directed towards the host galaxy group and not NW-SE as
we envisage here. Nevertheless, the direction of the gravitational
acceleration could plausibly be transverse to the jet axis if 
the separation (in line of sight distance) between the host and 
SW groups is closer than that inferred by attributing redshifts to
distances.  As noted in Sect.~\ref{s:local}, the velocity separation between the
two galaxy concentrations is 290 km s$^{-1}$. However, peculiar motions
could contribute significantly to this velocity difference, placing them closer in real 
space than one might infer from the redshifts alone. 
In this case,
the gravitational field in the vicinity of the SW lobe would be dominated
by both the host and SW galaxy groups, and we would expect the gravitational
acceleration vector to be directed between these two mass concentrations,
and roughly SE-NW. 
\par
\begin{figure*}
\centering
\includegraphics[angle=0, width=0.5\textwidth]{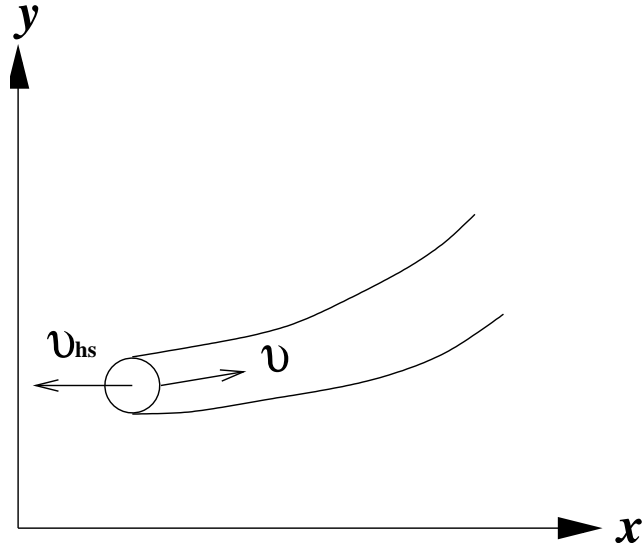}
\caption
{Schematic of the buoyant jet backflow in the $x$-$y$-plane.}
\label{f:curvature}
\end{figure*}

In the analysis below we assume that hydrostatic equilibrium defines the pressure 
and density distributions; heating and sweeping up of the ambient gas by the 
expanding radio lobe is neglected. A possible in-fall of the radio source 
and local environment  towards the centre of the large-scale filament is also 
ignored since peculiar velocities are unknown. We use the following 
notation: $\rho$ and $\rho_{\rm ext}$ are the mass densities of
the jet backflow and external medium respectively, $p$ and $p_{\rm ext}$ are 
the pressures in the lobe backflow and external medium respectively,
$v_{\rm bf}$ is the velocity of the backflow, $R_{\rm c}$ is the radius of curvature in the backflow, 
and  $g$ is the gravitational field which causes an hydrostatic pressure gradient 
in the ambient gas. We define a coordinate system such that the flow is in the $x$-$y$-plane, with the propagation of the jet hot spot along the $x$-axis, as visualized in
Fig.~\ref{f:curvature}. The velocity of the hot spot, $v_{\rm hs}$, and velocity of
the backflow, $v_{\rm bf}$, are both in the frame of the host galaxy.

Following \citet{worrall95} (and \citet{schatzmann78} in more detail), the dynamics 
of a buoyant jet backflow in such a medium are descirbed by Euler's  equation

\begin{equation}
\frac{\partial \bmath{v_{\rm bf}}}{\partial t} + (\bmath{v_{\rm bf}}\cdot\nabla)
\bmath{v_{\rm bf}} = -\frac{\nabla p}{\rho} +  \bmath{g},
\label{e:euler}
\end{equation}
where the gravitational field, the external density and pressure of the hydrostatic 
are related by:
\begin{equation}
\bmath{g} = \frac{\nabla p_{\rm ext}}{\rho_{\rm ext}}.
\label{e:grav}
\end{equation}
We assume that at each location along the backflowing lobe, the lobe plasma
is in pressure equilibrium with the external gas. Hence, the momentum equations
for the backflow become: 
\begin{equation}
\frac{\partial \bmath{v_{\rm bf}}}{\partial t} + (\bmath{v_{\rm bf}}
\cdot \nabla)\bmath{v_{\rm bf}} = 
- \nabla p_{\rm ext} \left( \frac{1}{\rho} - \frac{1}{\rho_{\rm ext}} \right).
\end{equation}
It therefore follows that the velocity, $v_{\rm bf}$, of the backflow and the radius of curvature,
$R_{\rm c}$, are related by

\begin{equation}
 \frac{v_{\rm bf}^2}{R_{\rm c}} =  
 g \, \frac{\rho_{\rm ext}}{\rho} \,
 \left( 1 - \frac {\rho}{\rho_{\rm ext}} \right).
 \label{e:v2}
\end{equation}
We therefore have:
\begin{equation}
 \frac{v_{\rm bf}^2}{R_{\rm c}} \approx \frac{\rho_{\rm ext}}{\rho}g
 \label{e:v2}
\end{equation}
for $\rho << \rho_{\rm ext}$. 
\par
Let us now consider the various parameters, which enter into
equation~(\ref{e:v2}), namely the velocity of the backflow, the density of the
external medium, and the density of the lobe
material. We have already estimated  $g$ in the vicinity of the radio
source (see Sect.~\ref{s:local}); based on the projected geometry of the SW lobe,
we estimate the backflowing SW
lobe plasma to have a radius of curvature of approximately 250 kpc.
\par
Recent estimates of the backflow velocity in giant radio galaxies have indicated
$v_{\rm bf} =$ 0.03 - 0.04$c$ \citep[e.g.][]{lara00,jamrozy05}. These estimates are
based on spectral aging techniques and rely on the argument, based on source
morphologies, that the mean backflow velocity is equal to the 
mean head advance velocity (i.e. $\langle v_{\rm bf} \rangle  = \langle v_{\rm hs} \rangle$).
In Sect.~\ref{s:hist}, we estimated that the time elapsed since the particles in the 
SW lobe extension were last accelerated is $>4\times10^{7}$ yr. 
These particles are separated from the current location of the jet termination by a 
projected distance of 630 kpc. Therefore, we estimate a separation velocity
(the lobe velocity with respect to the hot spot) of $<0.05c$. If we also assume equal 
advance and backflow velocities, then $v_{\rm bf} < 0.025c$; the upper limit is 
similar to the values derived for other giant radio sources.
\par
We estimate the external gas density using our knowledge of the 
3-D galaxy distribution. Our analysis suggests that the source \rg does not
reside in an extremely over-dense environment ({\it e.g.},  
a galaxy cluster); instead, we have shown that the host galaxy is a member of
a loose group that is embedded within a large-scale galaxy filament. 
The large-scale structure represents only moderate over-densities. Therefore,
in terms of the mean baryon density, $\rho_{b}$, we expect that 
$1< \rho_{\rm ext}/\rho_{b} < 100$ in the galaxy filament environment of the 
SW lobe.
\par
In order to estimate the density of the lobe material in
\rgns, we consider two possibilities: (1) that the lobe
consists purely of waste jet material, and (2) that in addition
to the waste jet material, the lobe is contaminated by thermal IGM, 
which has been entrained into the cocoon during the evolution of 
the radio source.
In estimating the density of the non-thermal waste jet material, 
we assume an electron-proton plasma with a power-law distribution in 
energy and index $a$. The lower and upper Lorentz factor cutoffs to this 
distribution are $\gamma_{\rm min}$ and $\gamma_{\rm max}$ respectively. 
In this case the internal mass density of the lobe plasma is related to the 
energy density, $u_{\rm p}$, in the synchrotron emitting particles by
\begin{equation}
\rho = \frac{m_{\rm p}}{m_{e} c^{2}} \frac{ (a-2)}{(a-1)}
\gamma_{\rm min}^{-1} u_{\rm p},
\label{e:rho}
\end{equation}
where we have assumed that $\gamma_{\rm max} \gg \gamma_{\rm min}$. 
We compute this expression, in conjunction with minimum energy
assumptions, using minimum Lorentz factors of  
$\gamma_{\rm min} = 10^{2}$ and $10^{3}$, 
since values within this range have been inferred for jets and lobes 
previously \citep[see e.g.][]{blundell06,worrall06}. For the particle energy density
calculation, we  adopt a maximum Lorentz factor of 
$\gamma_{\rm max}=10^5$.
For example, $\gamma_{\rm min} = 10^{2}$, gives $u_{\rm p}=4.5\times10^{-15}$ 
J m$^{-3}$ and  $\rho=2.8\times10^{-31}$ kg m$^{-3}$, while 
$\gamma_{\rm min} = 10^{3}$, results in $u_{\rm p} = 2.6\times10^{-15}$  J m$^{-3}$ and 
$\rho = 1.6\times10^{-32}$ kg m$^{-3}$. The implied lobe densities, expressed as a 
fraction of the external gas density, $\kappa=\rho/\rho_{\rm ext}$, are in the range
$10^{-7}<\kappa<10^{-4}$.

\par
For comparison, an electron-positron jet with the same $\gamma_{\rm min}$ values
would result in lobe densities that are a factor $\gtrsim10^{3}$ lower than those above
and a density ratio in the range $10^{-10}<\kappa<10^{-7}$.  For each plasma type, 
we note that the lobe densities and  corresponding $\kappa$-values are 
lower limits since they do not  account for possible entrainment of the relatively dense 
IGM.  
\par
Any entrainment of IGM gas into the cocoon of the
radio galaxy would pollute the lobe cavity and increase its density. 
This quantity, however, would remain less than that in the IGM
so that $\kappa<1$. The actual amount of entrainment might depend 
on the source history,
the power of the jets and the environment in which the source
is situated. We have examined the polarization properties over the bulk of the 
SW lobe, in an attempt to shed light on their thermal content, 
but find that these do not provide any useful limits on $\rho$. 
Rather, the measured polarization indicates that the Faraday depth 
over the source is $<1$ radian. This implies a thermal electron density of 
$<70$ m$^{-3}$ for the SW lobe, which has an estimated depth of 430 kpc 
and a  minimum energy magnetic field  0.1 nT.
If we allow for the presence of protons, the implied mass 
density is $<7\times10^{-26}$ kg m$^{-3}$. At the upper limit, this implies 
$\kappa>1$ for the considered range of $\rho_{\rm ext}$ values, 
and therefore, is a poor constraint.
\par

The inner sources in restarting or double-double radio galaxies
are a useful probe of the plasma in which they are enveloped. 
\citet{kaiser00} examined a sample of double-double restarting radio sources
(with sizes $\gtrsim1$~Mpc and in a IGM environment) and found 
$\kappa\sim0.001$ to be consistent with the observed properties of the 
inner doubles. More recently, \citet{safouris08} showed that entrainment 
has significantly increased the densities in the outer lobes of the giant 
double-double radio galaxy PKS B1545$-$321. They find a similar 
non-thermal density  in the lobes  to what we have inferred for \rgns; 
however, they also find that the \textit{total} lobe densities are much greater, 
and at least a few percent of the IGM density, because of contaminating thermal 
gas. We therefore consider that  $\kappa$ could be as high as 
0.01 if \rg has entrained IGM gas during its evolution. In such a case,
it is interesting to note that if the jets are electron-positron in composition, then
the lobe density would be completely dominated by the entrained gas.   
\par
For buoyancy to account for the movement of the SW lobe that has a
density $\rho=\kappa\rho_{\rm ext}$, the gravitational acceleration
is required to be

\begin{equation}
\label{e:kappa}
g = 7\times10^{-9} \kappa 
\left( \frac{v_{\rm bf}}{7500~\mbox{km s$^{-1}$}} \right)^{2} 
\left( \frac{R_{\rm c}}{250~\mbox{kpc}} \right)^{-1}
~\mbox{ m s$^{-2}$},
\end{equation}
for our adopted values of $R_{\rm c}$ and $v_{\rm bf}$. Thus for a lobe 
with $\kappa=0.01$, it is necessary that $g\approx 7\times10^{-11}$ m s$^{-2}$. 
For a lobe with no entrainment and $\kappa=10^{-4}$, the required 
acceleration is $g=7\times10^{-13}$ m s$^{-2}$.
\par

\begin{figure*}
\includegraphics[angle=0, width= 0.7\textwidth]{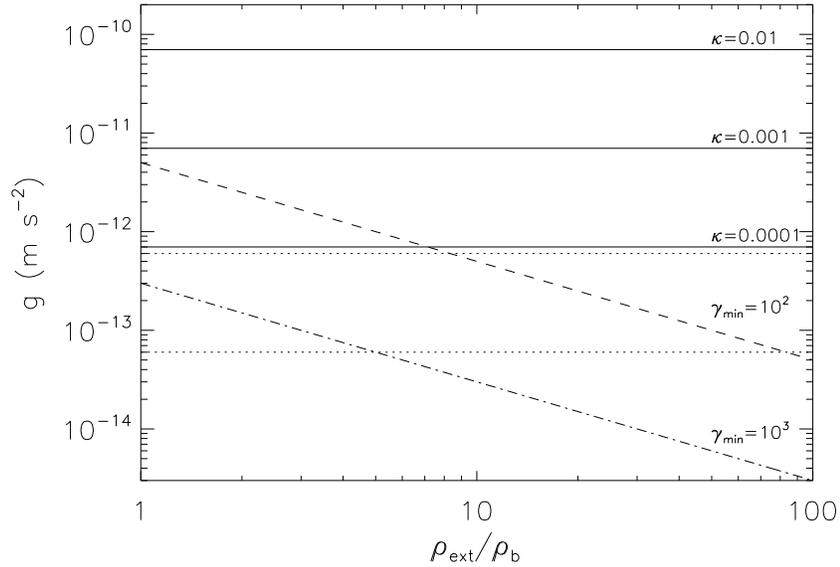}
\caption
{Gravitational acceleration as a function of external density, expressed 
in terms of the mean Baryon density in the local Universe. The 
accelerations required to move thermally contaminated lobes of density 
$\kappa \rho_{\rm ext}$ are shown with solid lines.
Also shown are the gravitational accelerations required to buoyantly
move a lobe with no thermal contamination and
$\gamma_{\rm min} = 10^{2}$ (dashed line) or $\gamma_{\rm min}=10^{3}$ 
(dot-dashed line). The lower and upper horizontal dotted lines show the 
 the estimated magnitudes of the gravitational field at the location of the SW lobe 
measured from local and large-scale galaxy distributions. }
\label{f:gz}
\end{figure*}

In Fig.~\ref{f:gz} we have plotted the value of the gravitational field that
is necessary to deflect the lobe backflow for a range of $\kappa$ values (solid lines). 
Estimates of the magnitude of the gravitational  field in the vicinity of the SW radio 
lobe based on the local and large-scale galaxy distributions are 
displayed with dotted lines. We also plot as a function of external gas density, 
the field that is required to buoyantly move the lobe if there has been no 
entrainment and  $\gamma_{\rm min} = 10^{2}$ (dashed line) or 
10$^{3}$ (dot-dashed line).
In this latter calculation, we have expressed the external density as a 
fraction of the mean baryon density of the Universe, $\rho_{b}$, 
which is the baryon density parameter $\Omega_{\rm Baryon}$ times the 
critical density $\rho_{\rm critical}$. 
\par
Figure~\ref{f:gz} shows that  if the amount of entrained thermal gas in the cocoon 
is negligible, and therefore the cocoon is relatively light ($10^{-7}<\kappa<10^{-4}$), 
the gravitational field required to deflect the lobe via buoyancy encompasses the 
range of what we estimate at the SW lobe 
($g\approx6\times10^{-14}$ - $6\times10^{-13}$ m s$^{-2}$), due to the surrounding mass distribution.
From the above analysis, we conclude that the off-axis evolution in the SW lobe 
may result from buoyancy forces in the backflow, if there is minimal 
entrainment. Alternatively, this may also occur if the entrained
gas is confined to a narrow boundary layer along the edge of the lobe.
The external densities implied by the buoyancy model are in good agreement
with those expected in the galaxy filament environment of the giant radio source. 
They represent moderate over-densities in the gas distribution and are 
therefore within those expected for the warm-hot phase of the IGM
\citep[e.g.][]{bregman07}. 
The external pressures signified by the model also hint at an ambient gas
with warm-hot temperatures. Pressure equilibrium
between the lowest surface brightness regions of the SW lobe and the 
external gas implies an IGM pressure of $1\times10^{-15}$~N~m$^{-2}$ 
in the vicinity of the flow. For the considered range of external densities,
this equates to an IGM temperature in the range $2\times(10^{6}-10^{8})$ K.
At the lower end (corresponding to higher external densities), the gas
temperature is  well matched to that of the WHIM (i.e. $10^{5}-10^{7}$ K). 
Therefore, a buoyant movement of the lobe plasma would support our initial 
hypothesis that  the lobes of \rg are interacting with the warm-hot gas phase
of the IGM. 
\par
The above interpretation is an appealing one since it signifies evidence
of an interaction between the radio source and WHIM. Nevertheless, we 
consider  below further mechanisms which may drive the off-axis movement 
of the SW lobe plasma.

\subsection{Buoyant bubbles}

We next consider a variant of the above model, where the SW lobe extension is treated
as a bubble rather than a back flow. The synchrotron bubble is embedded within
a thermal gaseous medium and rises against the gravity vector. 
In this case, the dynamics are determined by 
the balance between the buoyant and drag forces which act upon the bubble
in the IGM \citep[e.g.][]{gull73,bruggen01}. Equating these forces implies a terminal
velocity for the bubble of 
\begin{equation}
v_{\rm T} = \sqrt{\frac{2gV}{AC_{\rm D}}}
\end{equation}
where $V$ is the volume of the rising bubble, $A$ is the area of its cross-section,
and $C_{\rm D}$ is the drag coefficient. We approximate the displaced 
SW lobe as a sphere with $V/A = 130$~kpc and take $C_{\rm D} = 0.75$ 
\citep{churazov01}. Adopting a
gravitational field $g = 5.9\times10^{-13}$ m s$^{-2}$, as estimated at the location
of the host galaxy in the large-scale galaxy filament, we calculate a terminal
velocity $v_{\rm T} = 80$~km~s$^{-1}$ for the lobe. At this
speed, it would take more than $2\times10^{9}$ yr for the lobe plasma to move the 
observed 200 kpc distance off-axis. 
This time-scale is grossly longer than the estimated spectral ages ($\sim10^{7}$ yr)
for different components of the source ---including the extension at the 
NW end of the SW lobe 
---and signifies that buoyant forces are unlikely to be responsible for the movement of 
the lobe if it is a relict bubble. More generally, this analysis suggests that buoyant 
forces acting on bubbles of synchrotron lobe plasma in the IGM are unimportant unless 
the bubbles have significantly larger radii and are in gravitational fields that are at least
an order in magnitude or more greater than what we have estimated for \rgns; 
such gravitational fields are unlikely outside rich clusters
of galaxies.

\subsection{Density gradients in the IGM}

The most striking feature in the morphology of the giant radio source 
MRC B0319$-$454 is the side-to-side asymmetry in the lobe extents from the
core. The  NE lobe extends over a deprojected distance of 
640 kpc from the core, while the end of SW lobe is at a much greater
distance of 1430 kpc. Light-travel time effects 
are one possible mechanism for causing such 
asymmetries; however, as noted in \citet{saripalli94}, these effects would 
be small for the range of velocities that the 
ends of the lobes
might take. A better 
explanation relates to corresponding asymmetries in the IGM. In this section,
we examine the ambient gas distribution for such asymmetries. Again,
it is assumed that the galaxies trace the matter.
\par
The galaxy distribution within the host galaxy group suggests that
there might be a gas density gradient about the host. Specifically,
the host galaxy appears to lie at the center of the group. The most
luminous members of the group are located to the north of the host, with the
exception of the second brightest member, which resides west of the host.
Since low luminosity X-ray groups show evidence for irregular distributions
in the X-ray emitting intra-group gas, with the emission preferentially
biased to the luminous galaxies in the group  \citep{mulchaey00},
we might expect there to be less thermal gas south of the 
radio core in this group.  
\par
 
Based on the asymmetry in the lobe extents from the core, we infer that
the ends of the SW and NE lobes have advanced with a velocity ratio
of 2:1. Assuming that the advance of the leading lobe edge 
 is ram pressure limited  
 by the density of the external gas, we expect that the external density is 4
times larger on the NE side of the core compared to that on the SW side.
In Sect.~\ref{s:local}, we noted that there was evidence in the surrounding
galaxy distribution for a density gradient in the IGM along the radio 
jet axis of the source (see also Fig.~\ref{f:lods}). At a smoothing scale of 1.25 Mpc, the fractional 
over-density in the galaxy counts is about 10 in the region 
of the NE lobe and decreases along the axis of the southern jet to approximately 
2 in the vicinity of the SW hot spot. Assuming that the galaxies trace the gas, this 
gradient in the fractional density contrast, together with the derived mean
galaxy number density of 0.3, is consistent with the side-to-side length 
asymmetry in this source. Again, this suggests that the length
asymmetry in \rg results from density gradients in the IGM environment of the
source.  Also, the consistency between the derived ratio
of galaxy densities on each side of the core with the  
asymmetry in the lobe extents from the core,
supports our hypothesis that the galaxies are a reliable
tracer of the gas on large scales.
\par
Given that density gradients in the IGM can account for the side-to-side
length asymmetry in \rgns, it is worth considering if these
are also responsible for the observed asymmetry in the lobe shapes on the 
two sides of the core. We have shown previously that the NE lobe appears 
to be embedded within the loose group of galaxies of which the host is a 
member. The fractional over-density is relatively high ($\Delta n/\bar{n} \approx 10$) 
in the region of the NE lobe. Further examination of the fractional over-density 
in the vicinity of the radio source, at a smoothing scale of 1.25 Mpc, shows no 
evidence for density gradients within the region of the group occupied by the 
NE lobe. Slice profiles along the length of the NE lobe, along the line of sight 
through the NE lobe, and transverse to the axis of the NE lobe, show that the 
fractional overdensities are a constant and a little over 10 in each of these 
directions. The fractional over-density values do decline at the periphery of 
the group. However, the NE lobe appears to be located close to the centre of 
the galaxy density distribution associated with the group.
The uniform density in the region occupied by the NE lobe is consistent 
with the axially symmetric backflow of the hot spot plasma along the radio axis 
in this lobe. The accompanying high fractional over-density values in the 
region of the galaxy group, along with the radio morphology of the
source, suggest that the relatively denser intra-group medium confines
the NE lobe and stalls any expansion losses. This has resulted in a relatively 
bright and luminous NE lobe. 
\par
A similar analysis of the fractional over-density values in the region of the 
SW lobe suggests a different scenario. As mentioned
already, there is a decline in the fractional over-density along the jet axis
from the radio core to the lobe end. In addition, we find that there is a gradient in the 
over-density along a line of sight through the SW lobe:
the fractional over-density is about a factor of 2 larger behind the lobe
compared to that in front, and at the location of the lobe. Also,
as might be expected from the radio morphology, there exists a gradient 
in the over-density that is transverse to the radio jet axis and oriented SE to NW. 
Fractional over-density values along this direction
show that there is more galaxy density on the SE side
of the source compared to the NW side. This gradient may be viewed as arising
from the combined effects of the host group and neighbouring
galaxy concentration. The numbers we derive suggest that the external 
densities are a factor of 2 or more higher on the SE side of the lobe
relative to the NW side. 
\par
Such a density gradient may have caused the  asymmetric 
expansion of the SW lobe plasma in this direction. 
The plasma on the NW side of the radio jet axis extends about
three times further from the jet axis, than the plasma on the
SE side. The plasma may be expanding on the two sides with 
a velocity ratio of 3:1. 
If this side-ways expansion is ram pressure limited, 
then we expect the external density on the SE side of the axis to be a factor 
9 larger than that on the NW side. 
Our analysis of the fractional density contrast indicates that the gas densities
on the two sides of the SW lobe are in the ratio 2:1 when examined
with a smoothing scale 1.25~Mpc; but these could be larger on 
smaller smoothing scales. Although the density
gradient is in the direction expected for this model, the local
galaxy number density ($<0.3$ at this smoothing scale) is 
insufficient to make a useful estimate of density
variations on the 0.5~Mpc length scale, which is necessary 
to examine whether the magnitude of the density
gradient is sufficient to account for the transverse expansion.
\par
In particular,  the decreased ambient density
 to the NW of the SW lobe, may have caused the synchrotron plasma to break
 confinement along this edge. In this case, the low surface brightness
 extension and its relatively steep spectral index distribution, could
 be attributed to expansion losses, provided that the electron energy
 spectrum is curved,  or has a high frequency break that has shifted to the
 observing range. 

\par
Finally, we note that the extent of the NE lobe from the core is similar 
to the extent of the detected SW jet from the core.  The gas associated 
with the group may be distributed to equal extents about the host
(although as noted earlier, we expect more gas to the north based on the
locations of the most luminous galaxies).
The gaseous envelope, which likely confines the NE lobe, might
also be responsible for the confinement and detectability of the SW jet 
in this region.
\par
To summarize, the qualitative correspondance between the density 
distribution and the radio source morphology is suggestive of a physical 
origin for the structure and asymmetries in the interaction between the 
radio jet plasma
 and ambient gas inhomogenieties: a likely mechanism is ram 
pressure limitation of the lobe advance and expansion
by the gas density.
In this picture, the asymmetries are caused by the locations and orientations 
of the radio lobes with respect to the gaseous media associated with the host 
galaxy group and the neighbouring galaxy concentration south of the host. 

\section{Conclusions}
\label{s:con}

We have presented new high sensitivity radio images of the powerful
radio galaxy \rg together with observations of the surrounding 
large-scale structure.
Using these we have sought to understand the influence of the ambient 
gas and gravity on the evolution of  the jets and post hot spot plasma in 
this remarkable source. Our observations are qualitatively consistent 
with a model wherein the galaxies trace inhomogeneities in the
intergalactic gas, and these in turn determine the evolution of the 
giant radio source.  
In this respect, our study has reached similar conclusions to those of 
\citet{subrahmanyan08}, who also showed that asymmetries in the morphology
of the giant relict source, MSH J0505$-$2835, may
be governed by inhomogeneities in the ambient intergalactic gas distribution,
whose distribution follows the large scale structure in galaxies. 
In particular, both studies have found a foreshortening 
of the lobes in the direction of greater galaxy density, and off-axis
distortions in the direction of decreasing galaxy density. 

\par
In \rgns, 
over-densities in the galaxy distribution are found north of 
the radio core. These are related to the loose galaxy group of which the
host is a member. The advance of the NE jet and expansion of the 
corresponding lobe are ram-pressure limited
by the intra-group medium of this galaxy concentration.
Furthermore, there is less gas density to the south of the radio core, 
explaining the rapid advance of the SW jet and increased expansion
losses in the post hot spot flow. A density gradient, perpendicular to the jet
axis in the SW lobe, may also influence the asymmetric lobe expansion 
to the NW.
\par
We further hypothesize that the SW backflow may be moving
away from the source axis and NW due to buoyancy.
We find that buoyant forces (due to a gravitational 
field transverse to the radio source axis) can deflect 
the backflow of the SW jet plasma, provided there has been
minimal entrainment of ambient gas into the lobe. 
This leads to a model in which the medium external
to the SW radio lobe has the same properties as those
inferred for the WHIM component of the IGM.
\par
Future deep imaging of the gas associated with the large-scale structure 
would help to further our interpretation. This work
demonstrates the usefulness of giant radio galaxies as probes
of the IGM through the relationship between source morphology
and the ambient large-scale structure distribution.

\section{Acknowledgments}
The Australia Telescope Compact Array is part of the Australia Telescope, which is funded by the Commonwealth of Australia for operation as a National Facility managed by CSIRO. This research has made use of the NASA/IPAC Extragalactic Database, which is operated by the Jet Propulsion Laboratory, California Institute of Technology, under contract with the National Aeronautics and Space Administration. We thank the Anglo-Australian Observatory for data obtained with the 2dF during service time and the AAOmega instrument during the Science
Verification program. We acknowledge the use of the RUNZ code written by Will J. Sutherland. We are grateful to Professor R. Ekers for useful discussions and comments on this manuscript.
We thank the anonymous referee for helpful comments and suggestions.




\begin{thebibliography}{}

\bibitem[\protect\citeauthoryear{{Begelman}, {Blandford} \& {Rees}}{{Begelman}
  et~al.}{1984}]{begelman84}
{Begelman} M.~C.,  {Blandford} R.~D.,    {Rees} M.~J.,  1984, Reviews of Modern
  Physics, 56, 255

\bibitem[\protect\citeauthoryear{{B{\^i}rzan}, {Rafferty}, {McNamara}, {Wise}
  \& {Nulsen}}{{B{\^i}rzan} et~al.}{2004}]{birzan04}
{B{\^i}rzan} L.,  {Rafferty} D.~A.,  {McNamara} B.~R.,  {Wise} M.~W.,
  {Nulsen} P.~E.~J.,  2004, \apj, 607, 800

\bibitem[\protect\citeauthoryear{{Bliton}, {Rizza}, {Burns}, {Owen} \&
  {Ledlow}}{{Bliton} et~al.}{1998}]{bliton98}
{Bliton} M.,  {Rizza} E.,  {Burns} J.~O.,  {Owen} F.~N.,    {Ledlow} M.~J.,
  1998, \mnras, 301, 609

\bibitem[\protect\citeauthoryear{{Blundell}, {Fabian}, {Crawford}, {Erlund} \&
  {Celotti}}{{Blundell} et~al.}{2006}]{blundell06}
{Blundell} K.~M.,  {Fabian} A.~C.,  {Crawford} C.~S.,  {Erlund} M.~C.,
  {Celotti} A.,  2006, \apjl, 644, L13

\bibitem[\protect\citeauthoryear{{Boehringer}, {Voges}, {Fabian}, {Edge} \&
  {Neumann}}{{Boehringer} et~al.}{1993}]{boehringer93}
{Boehringer} H.,  {Voges} W.,  {Fabian} A.~C.,  {Edge} A.~C.,    {Neumann}
  D.~M.,  1993, \mnras, 264, L25

\bibitem[\protect\citeauthoryear{{Bregman}}{{Bregman}}{2007}]{bregman07}
{Bregman} J.~N.,  2007, \araa, 45, 221

\bibitem[\protect\citeauthoryear{{Br{\"u}ggen} \& {Kaiser}}{{Br{\"u}ggen} \&
  {Kaiser}}{2001}]{bruggen01}
{Br{\"u}ggen} M.,  {Kaiser} C.~R.,  2001, \mnras, 325, 676

\bibitem[\protect\citeauthoryear{{Bryant} \& {Hunstead}}{{Bryant} \&
  {Hunstead}}{2000}]{Bryant00}
{Bryant} J.~J.,  {Hunstead} R.~W.,  2000, \apj, 545, 216

\bibitem[\protect\citeauthoryear{{Cen} \& {Ostriker}}{{Cen} \&
  {Ostriker}}{1999}]{cen99}
{Cen} R.,  {Ostriker} J.~P.,  1999, \apj, 514, 1

\bibitem[\protect\citeauthoryear{{Cen} \& {Ostriker}}{{Cen} \&
  {Ostriker}}{2006}]{cen06}
{Cen} R.,  {Ostriker} J.~P.,  2006, \apj, 650, 560

\bibitem[\protect\citeauthoryear{{Churazov}, {Br{\"u}ggen}, {Kaiser},
  {B{\"o}hringer} \& {Forman}}{{Churazov} et~al.}{2001}]{churazov01}
{Churazov} E.,  {Br{\"u}ggen} M.,  {Kaiser} C.~R.,  {B{\"o}hringer} H.,
  {Forman} W.,  2001, \apj, 554, 261

\bibitem[\protect\citeauthoryear{{Dav{\'e}}, {Cen}, {Ostriker}, {Bryan},
  {Hernquist}, {Katz}, {Weinberg}, {Norman} \& {O'Shea}}{{Dav{\'e}}
  et~al.}{2001}]{dave01}
{Dav{\'e}} R.,  {Cen} R.,  {Ostriker} J.~P.,  {Bryan} G.~L.,  {Hernquist} L.,
  {Katz} N.,  {Weinberg} D.~H.,  {Norman} M.~L.,    {O'Shea} B.,  2001, \apj,
  552, 473

\bibitem[\protect\citeauthoryear{{Douglass}, {Blanton}, {Clarke}, {Sarazin} \&
  {Wise}}{{Douglass} et~al.}{2008}]{douglass08}
{Douglass} E.~M.,  {Blanton} E.~L.,  {Clarke} T.~E.,  {Sarazin} C.~L.,
  {Wise} M.,  2008, \apj, 673, 763

\bibitem[\protect\citeauthoryear{{Fleenor}, {Rose}, {Christiansen}, {Hunstead},
  {Johnston-Hollitt}, {Drinkwater} \& {Saunders}}{{Fleenor}
  et~al.}{2005}]{fleenor05}
{Fleenor} M.~C.,  {Rose} J.~A.,  {Christiansen} W.~A.,  {Hunstead} R.~W.,
  {Johnston-Hollitt} M.,  {Drinkwater} M.~J.,    {Saunders} W.,  2005, \aj,
  130, 957

\bibitem[\protect\citeauthoryear{{Fleenor}, {Rose}, {Christiansen},
  {Johnston-Hollitt}, {Hunstead}, {Drinkwater} \& {Saunders}}{{Fleenor}
  et~al.}{2006}]{fleenor06}
{Fleenor} M.~C.,  {Rose} J.~A.,  {Christiansen} W.~A.,  {Johnston-Hollitt} M.,
  {Hunstead} R.~W.,  {Drinkwater} M.~J.,    {Saunders} W.,  2006, \aj, 131,
  1280

\bibitem[\protect\citeauthoryear{{Gull} \& {Northover}}{{Gull} \&
  {Northover}}{1973}]{gull73}
{Gull} S.~F.,  {Northover} K.~J.~E.,  1973, \nat, 244, 80

\bibitem[\protect\citeauthoryear{{Hardcastle}, {Evans} \&
  {Croston}}{{Hardcastle} et~al.}{2007}]{hardcastle07}
{Hardcastle} M.~J.,  {Evans} D.~A.,    {Croston} J.~H.,  2007, \mnras, 376,
  1849

\bibitem[\protect\citeauthoryear{{Hardcastle}, {Sakelliou} \&
  {Worrall}}{{Hardcastle} et~al.}{2005}]{hardcastle05}
{Hardcastle} M.~J.,  {Sakelliou} I.,    {Worrall} D.~M.,  2005, \mnras, 359,
  1007

\bibitem[\protect\citeauthoryear{{Haslam}, {Salter}, {Stoffel} \&
  {Wilson}}{{Haslam} et~al.}{1982}]{haslam82}
{Haslam} C.~G.~T.,  {Salter} C.~J.,  {Stoffel} H.,    {Wilson} W.~E.,  1982,
  \aaps, 47, 1

\bibitem[\protect\citeauthoryear{{Haverkorn}, {Katgert} \& {de
  Bruyn}}{{Haverkorn} et~al.}{2000}]{haverkorn00}
{Haverkorn} M.,  {Katgert} P.,    {de Bruyn} A.~G.,  2000, \aap, 356, L13

\bibitem[\protect\citeauthoryear{{Huchra} \& {Geller}}{{Huchra} \&
  {Geller}}{1982}]{huchra82}
{Huchra} J.~P.,  {Geller} M.~J.,  1982, \apj, 257, 423

\bibitem[\protect\citeauthoryear{{Ishwara-Chandra} \&
  {Saikia}}{{Ishwara-Chandra} \& {Saikia}}{1999}]{ishwara99}
{Ishwara-Chandra} C.~H.,  {Saikia} D.~J.,  1999, \mnras, 309, 100

\bibitem[\protect\citeauthoryear{{Jamrozy}, {Klein}, {Machalski} \&
  {Mack}}{{Jamrozy} et~al.}{2004}]{jamrozy04}
{Jamrozy} M.,  {Klein} U.,  {Machalski} J.,    {Mack} K.-H.,  2004, in
  {M{\'u}jica} R.,  {Maiolino} R.,  eds, Multiwavelength AGN Surveys
  {Large-Scale Radio Structure in the Universe: Giant Radio Galaxies}.
p.~431

\bibitem[\protect\citeauthoryear{{Jamrozy}, {Machalski}, {Mack} \&
  {Klein}}{{Jamrozy} et~al.}{2005}]{jamrozy05}
{Jamrozy} M.,  {Machalski} J.,  {Mack} K.-H.,    {Klein} U.,  2005, \aap, 433,
  467

\bibitem[\protect\citeauthoryear{{Jones}, {Saunders}, {Colless}, {Read},
  {Parker}, {Watson}, {Campbell}, {Burkey}, {Mauch} \& {The 6dFGS
  team}}{{Jones} et~al.}{2004}]{jones04a}
{Jones} D.~H.,  {Saunders} W.,  {Colless} M.,  {Read} M.~A.,  {Parker} Q.~A.,
  {Watson} F.~G.,  {Campbell} L.~A.,  {Burkey} D.,  {Mauch} T.,    {The 6dFGS
  team} 2004, \mnras, 355, 747

\bibitem[\protect\citeauthoryear{{Jones}, {Saunders}, {Read} \&
  {Colless}}{{Jones} et~al.}{2005}]{jones05a}
{Jones} D.~H.,  {Saunders} W.,  {Read} M.,    {Colless} M.,  2005, Publications
  of the Astronomical Society of Australia, 22, 277

\bibitem[\protect\citeauthoryear{{Jones}}{{Jones}}{1989}]{jones89}
{Jones} P.~A.,  1989, Proceedings of the Astronomical Society of Australia, 8,
  81

\bibitem[\protect\citeauthoryear{{Kaiser}, {Schoenmakers} \&
  {R{\"o}ttgering}}{{Kaiser} et~al.}{2000}]{kaiser00}
{Kaiser} C.~R.,  {Schoenmakers} A.~P.,    {R{\"o}ttgering} H.~J.~A.,  2000,
  \mnras, 315, 381

\bibitem[\protect\citeauthoryear{{Kaiser}}{{Kaiser}}{1987}]{kaiser87}
{Kaiser} N.,  1987, \mnras, 227, 1

\bibitem[\protect\citeauthoryear{{Klamer}, {Subrahmanyan} \&
  {Hunstead}}{{Klamer} et~al.}{2004}]{klam04}
{Klamer} I.,  {Subrahmanyan} R.,    {Hunstead} R.~W.,  2004, \mnras, 351, 101

\bibitem[\protect\citeauthoryear{{Kraft}, {Hardcastle}, {Worrall} \&
  {Murray}}{{Kraft} et~al.}{2005}]{kraft05}
{Kraft} R.~P.,  {Hardcastle} M.~J.,  {Worrall} D.~M.,    {Murray} S.~S.,  2005,
  \apj, 622, 149

\bibitem[\protect\citeauthoryear{{Lara}, {Mack}, {Lacy}, {Klein}, {Cotton},
  {Feretti}, {Giovannini} \& {Murgia}}{{Lara} et~al.}{2000}]{lara00}
{Lara} L.,  {Mack} K.-H.,  {Lacy} M.,  {Klein} U.,  {Cotton} W.~D.,  {Feretti}
  L.,  {Giovannini} G.,    {Murgia} M.,  2000, \aap, 356, 63

\bibitem[\protect\citeauthoryear{{Ledlow} \& {Owen}}{{Ledlow} \&
  {Owen}}{1996}]{ledlow96}
{Ledlow} M.~J.,  {Owen} F.~N.,  1996, \aj, 112, 9

\bibitem[\protect\citeauthoryear{{Lewis}, {Cannon}, {Taylor}, {Glazebrook},
  {Bailey}, {Baldry}, {Barton}, {Bridges} \& {Dalton}}{{Lewis}
  et~al.}{2002}]{lewis02}
{Lewis} I.~J.,  {Cannon} R.~D.,  {Taylor} K.,  {Glazebrook} K.,  {Bailey}
  J.~A.,  {Baldry} I.~K.,  {Barton} J.~R.,  {Bridges} T.~J.,    {Dalton}
  G.~B.~e.,  2002, \mnras, 333, 279

\bibitem[\protect\citeauthoryear{{McNamara}, {Wise}, {Nulsen}, {David},
  {Sarazin}, {Bautz}, {Markevitch}, {Vikhlinin}, {Forman}, {Jones} \&
  {Harris}}{{McNamara} et~al.}{2000}]{mcnamara00}
{McNamara} B.~R.,  {Wise} M.,  {Nulsen} P.~E.~J.,  {David} L.~P.,  {Sarazin}
  C.~L.,  {Bautz} M.,  {Markevitch} M.,  {Vikhlinin} A.,  {Forman} W.~R.,
  {Jones} C.,    {Harris} D.~E.,  2000, \apjl, 534, L135

\bibitem[\protect\citeauthoryear{{Mills}, {Slee} \& {Hill}}{{Mills}
  et~al.}{1960}]{mills60}
{Mills} B.~Y.,  {Slee} O.~B.,    {Hill} E.~R.,  1960, Australian Journal of
  Physics, 13, 676

\bibitem[\protect\citeauthoryear{{Mulchaey}}{{Mulchaey}}{2000}]{mulchaey00}
{Mulchaey} J.~S.,  2000, \araa, 38, 289

\bibitem[\protect\citeauthoryear{{Nulsen}, {David}, {McNamara}, {Jones},
  {Forman} \& {Wise}}{{Nulsen} et~al.}{2002}]{nulsen02}
{Nulsen} P.~E.~J.,  {David} L.~P.,  {McNamara} B.~R.,  {Jones} C.,  {Forman}
  W.~R.,    {Wise} M.,  2002, \apj, 568, 163

\bibitem[\protect\citeauthoryear{{Safouris}, {Subrahmanyan}, {Bicknell} \&
  {Saripalli}}{{Safouris} et~al.}{2008}]{safouris08}
{Safouris} V.,  {Subrahmanyan} R.,  {Bicknell} G.~V.,    {Saripalli} L.,  2008,
  \mnras, 385, 2117

\bibitem[\protect\citeauthoryear{{Saripalli}, {Subrahmanyan} \&
  {Hunstead}}{{Saripalli} et~al.}{1994}]{saripalli94}
{Saripalli} L.,  {Subrahmanyan} R.,    {Hunstead} R.~W.,  1994, \mnras, 269, 37

\bibitem[\protect\citeauthoryear{{Schatzmann}}{{Schatzmann}}{1978}]{schatzmann%
78}
{Schatzmann} M.,  1978, Zeitschrift Angewandte Mathematik und Physik, 29, 608

\bibitem[\protect\citeauthoryear{{Scheuer}}{{Scheuer}}{1974}]{scheuer74}
{Scheuer} P.~A.~G.,  1974, \mnras, 166, 513

\bibitem[\protect\citeauthoryear{{Simard-Normandin} \&
  {Kronberg}}{{Simard-Normandin} \& {Kronberg}}{1980}]{simard80}
{Simard-Normandin} M.,  {Kronberg} P.~P.,  1980, \apj, 242, 74

\bibitem[\protect\citeauthoryear{{Slee}}{{Slee}}{1995}]{slee95}
{Slee} O.~B.,  1995, Australian Journal of Physics, 48, 143

\bibitem[\protect\citeauthoryear{{Subrahmanyan}, {Saripalli}, {Safouris} \&
  {Hunstead}}{{Subrahmanyan} et~al.}{2008}]{subrahmanyan08}
{Subrahmanyan} R.,  {Saripalli} L.,  {Safouris} V.,    {Hunstead} R.~W.,  2008,
  \apj, 677, 63

\bibitem[\protect\citeauthoryear{{Worrall} \& {Birkinshaw}}{{Worrall} \&
  {Birkinshaw}}{2006}]{worrall06}
{Worrall} D.~M.,  {Birkinshaw} M.,  2006, in {Alloin} D.,  ed., Physics of
  Active Galactic Nuclei at all Scales Vol.~693 of Lecture Notes in Physics,
  Berlin Springer Verlag, {Multiwavelength Evidence of the Physical Processes
  in Radio Jets}.
p.~39

\bibitem[\protect\citeauthoryear{{Worrall}, {Birkinshaw} \&
  {Cameron}}{{Worrall} et~al.}{1995}]{worrall95}
{Worrall} D.~M.,  {Birkinshaw} M.,    {Cameron} R.~A.,  1995, \apj, 449, 93

\bibitem[\protect\citeauthoryear{{Wright}, {Griffith}, {Burke} \&
  {Ekers}}{{Wright} et~al.}{1994}]{wright94}
{Wright} A.~E.,  {Griffith} M.~R.,  {Burke} B.~F.,    {Ekers} R.~D.,  1994,
  \apjs, 91, 111

\end{thebibliography}

\newpage
\begin{table*}\begin{center}
\caption{Journal of ATCA observations}
\label{t:journal}
\vspace{6pt}
\begin{tabular}{c c c c}\hline\hline \vspace{-0.1cm}   \\   
Array       & Frequencies  &  Date & Duration  \\              
&        (MHz)       &          & (hr) \\  \vspace{-0.1cm}   \\   
 \hline \vspace{-0.1cm}   \\   
EW367 &    1378/2368  &  2003 Sep 02 &  12 \\
750B&      1378/2368& 2003 Sep 18  &  12 \\
EW352&     1378/2368&  2003 Oct 02  &  12   \\
1.5D &     1378/2368&  2003 Nov 16  & 12  \\
1.5A &     1378/2368&  2004 Apr 02  & 12 \\
\vspace{-0.1cm}   \\   
 \hline \end{tabular}\end{center}\end{table*}

\begin{table*}
\begin{center}
\caption{Radio flux densities of \rgns.}
\label{t:fluxes}
\vspace{6pt}
\begin{tabular}{c c c c c c}
\hline \hline
\vspace{-0.1cm}   \\   
Frequency (MHz)     & Survey/         &  N lobe     & S lobe     & Total    &
Reference \\
    (MHz)                     &  Telescope   &    (Jy)       &     (Jy)     &
(Jy)    &                    \\   
\vspace{-0.1cm} \\          
\hline
\vspace{-0.1cm} \\     
80                 &  Culgoora            & 22            &            &
& \citet{slee95} \\
85.5                & MSH            &19            &            &      &
\citet{mills60} \\
160                & Culgoora        & 11.6        &            &      &
\citet{slee95} \\
408                & 408 MHz         &        &            & $10.8\pm2.1$&
\citet{haslam82} \\
                                    &  all-sky survey&               &
&                          &                          \\  843
& MOST            & 4.64        & 0.71        & $5.35\pm0.5$&
Saripalli et al. (1994) \\
1378                & ATCA            & 3.16        & 0.70        &
$3.86\pm0.2$& This paper \\
2368                & ATCA            & 1.99        & 0.47        &
$2.46\pm0.1$& This paper \\
4850                & PMN            & 1.04        & 0.27        &
$1.31\pm$0.06& \citet{wright94}\\
\vspace{-0.1cm} \\    
\hline 
\end{tabular}
\end{center}
\end{table*}


\bsp

\label{lastpage}

\end{document}